\documentstyle[eqsecnum,aps,pre,epsf]{revtex}
\begin{document}
\draft
\title{Transport Phenomena at a Critical Point \protect \\
-- Thermal Conduction in the Creutz Cellular Automaton --}
\author{K. Saito}
\address{
Department of Physics, Graduate School of Science, \\
Osaka University,
Toyonaka 560--0043, Japan}
\author{S. Takesue}
\address{
Faculty of Integrated Human Studies,
Kyoto University, \\
Kyoto 606--8501, Japan}
\author{S. Miyashita}
\address{
Department of Earth and Space Science, Faculty of Science,\\
Osaka University, Toyonaka 560--0043, Japan}
\date{\today}

\maketitle

\begin{abstract}
Nature of energy transport around a critical point is studied 
in the Creutz cellular automaton.  Fourier heat law is confirmed to hold
in this model by a direct measurement of heat flow under a temperature
gradient.  The thermal conductivity is carefully investigated near the 
phase transition by the use of the Kubo formula.  As the result,
the thermal conductivity is found to take a finite value at the 
critical point contrary to some previous works.
Equal-time correlation of the heat flow is also analyzed by
a mean-field type approximation to investigate the temperature 
dependence of thermal conductivity.  A variant of the Creutz cellular
automaton called the Q2R is also investigated and similar results are
obtained. 
\end{abstract}

\section{Introduction}

Creutz devised a deterministic dynamics for the two-dimensional
Ising model with a momentum term \cite{CR85}.  
This dynamics is a kind of cellular automaton (CA), where states 
are upadated in a deterministic way with energy conservation
and we call it the Creutz cellular automaton (CCA).  
In the CCA random numbers are not necessary for its time evolution,
which provides an advantage in numerical simulations.
Thus, the CCA and its variants have been used 
to investigate equilibrium properties of magnetic systems \cite{BK97}
instead of conventional Monte Carlo method, 
especially the critical phenomena at the critical point.

Besides this advantage, the CCA provides a dynamical model for time
evolution with energy conservation.
Thus the CCA can be used to study transport phenomena where 
flows of physical quantities take important roles.
In fact, numerical results for heat conduction in the CCA were
reported in \cite{CR85} and the thermal conductivity was found to
be proportional to $T^{-2}$ in high temperature, where $T$ denotes 
the temperature.
Harris and Grant showed that this temperature dependence is 
explained by the Kubo formula \cite{HG88}. 
They presented an asymptotic expression for the thermal conductivity
in the high and low temperature limits by evaluating the first term in 
the Kubo formula.

Then, it is natural to ask a possible connection between the thermal
conductivity and the phase transition.  Because the specific heat
diverges at the critical point of the Ising model, the thermal 
conductivity may also show some peculiarity at the point.  Actually,
in some materials, abnormal behavior of the thermal conductivity has 
been observed\cite{MMZPS}. 
Clearly, the CCA is suitable to look into the thermal conductivity
near the critical point.
In the above mentioned paper,
Harris and Grant made a comment that the thermal conductivity must
vanish at the critical temperature $T_{\rm C}$ without any evidences.
It is a purpose of our paper to clarify what really happens to the energy
transport at the critical point in the CCA.

Thus, in this paper we investigate the temperature dependence of the 
thermal conductivity in the thermodynamic limit.
We obtain the thermal conductivity by two methods.  
One is a direct measurement of heat flow under a temperature gradient.  
The validity of the Fourier heat law is established in a wide range of
temperature values and the coefficient of thermal conductivity is estimated.
The other is the use of the Kubo formula.  Explicit derivation
of the formula is given and the coefficient of thermal conductivity is 
calculated from equilibrium autocorrelations of the energy flow.
We check that both the methods yield the same result and 
find a finite conductivity at $T_{\rm C}$, 
which does not agree with the previous belief.

We also develop a mean-field approximation for the equal-time
correlation of the energy flow, which improves the estimate by Harris
and Grant\cite{HG88}.
Since it is the first term of the Kubo formula,
the result of this treatment not only explains the temperature 
dependence at the high and low temperature limits, but also gives 
a qualitatively good estimate for the overall temperature dependence.

The conditions under which the Fourier heat law is 
satisfied have been studied in the literature 
\cite{MB97,PR92,LLP97,Hatano,T90} mainly by using Hamiltonian 
systems. The dynamical rules of CA are so simple and local 
that fast simulations  are possible.  Thus,
one of the present authors (ST) applied CA to this 
problem and found some rules that clearly satisfy the Fourier 
heat law\cite{T90}.  However, most of the studies have so far been 
confined in one-dimensional models, which might cause pathological 
effects due to a single path of the flow.  
Here we study a two dimensional system with CCA where we are free 
from the above anxiety.
We find that the Fourier heat law holds at all temperatures in the CCA.
We also find that the Q2R, which is a variant of 
the CCR without momentum terms, satisfies the Fourier heat law 
in two dimensions, although energy transport in Q2R is ballistic
in one dimension.

This paper is organized as follows.  In Sec.\@ 2, our model and method
are explained and an expression for the local energy flux is
derived via the equation of continuity.
In Sec.\@ 3, we demonstrate that 
the Fourier heat law holds in a wide range of temperatures by a direct
simulation.
In Sec.\@ 4, the thermal conductivity is calculated by the use of the Kubo
formula and its temperature dependence is carefully investigated 
especially around $T_{\rm C}$.
A mean field analysis is done for the correlation of the
energy flux in Sec.\@ 5. 
Numerical results for the Q2R are exhibited in Sec.\@ 6. 
We give summary and discussion in Sec.\@ 7.

\section{Model}

The CCA is defined as follows.
Let us consider the square lattice.
A couple of variables
$(\sigma_{i,j},\tilde{\sigma}_{i,j})$ are assigned at a site $(i,j)$.
Here $\sigma_{i,j}\in\{+1, -1\}$ denotes a spin and 
$\tilde{\sigma}_{i,j}\in\{0,1,2,3\}$ is called a momentum.
Then the total Hamiltonian is given by 
\begin{equation}
  H = - \sum_{i,j} \left ( \sigma_{i,j}\sigma_{i+1,j} +
         \sigma_{i,j}\sigma_{i,j+1} \right ) 
+\sum_{i,j}4 \tilde{\sigma}_{i,j}.
\label{QRH}
\end{equation}
The first term represents the ferromagnetic Ising interaction 
between the nearest-neighbor spins and the second term 
represents a kind of {\em kinetic} energy.
We divide the lattice into two sublattices like the checkerboard. 
Site $(i,j)$ is called even or odd according as the sum $i+j$ is even 
or odd.
One unit of time evolution consists of 
two processes each of which simultaneously updates
variables on a sublattice.  Namely, when the variables are updated from the
states at time $t$, first the even sites are updated at
time $t+1/2$ and next the odd sites are updated at time $t+1$.  
The updating rule is the following.
Spin flip is accepted when the momentum at the site can compensate 
the energy change of the flip.
That is, if the following relation is satisfied, 
$0\le \tilde{\sigma}_{i,j}-\frac{1}{2}\sigma_{i,j}
\sum_{\rm nn}\sigma_{\rm nn}\le 3$,
where nn denotes the nearest neighbor sites of $(i,j)$,
the spin $\sigma_{i,j}$ changes its sign and the momentum is changed 
to conserve the total energy. 


Now we derive expressions for a local energy and an energy flux.
From the total Hamiltonian 
(\ref{QRH}), we can define the local energy on the site $(i,j)$ 
at time $t$ as
\begin{equation}
E_{i, j}^t = -\sigma_{i, j}(\sigma_{i+1,j} + \sigma_{i,j+1}) 
+ 4\tilde{\sigma}_{i, j}.
\label{LSH}
\end{equation}
Note that the total
energy is equal to the sum of the local energies over the lattice.
First we consider the case where the site $(i,j)$ is even.
If the spin at site $(i,j)$ is flipped at time $t+1/2$,
we have
\begin{eqnarray*}
\sigma_{i,j}^{t+1/2} &=& - \sigma_{i,j}^{t} , \\ 
\tilde{\sigma}_{i,j}^{t+1/2} &=& 
\tilde{\sigma}_{i,j}^{t} -\frac{1}{2}\sigma_{i,j}^{t}(\sigma_{i-1,j}^{t} 
+ \sigma_{i+1,j}^{t} + \sigma_{i,j-1}^{t} + \sigma_{i,j+1}^{t}) , \\
\end{eqnarray*}
and the difference between the local energies at times $t+1/2$ and $t$
is given by
\begin{eqnarray*}
E_{i,j}^{t+1/2} - E_{i,j}^{t} 
&=& -2 \sigma_{i,j}^{t}(\sigma_{i-1,j}^{t}+\sigma_{i,j-1}^{t}).  
\end{eqnarray*}
If the spin $\sigma_{i,j}$ is not flipped, the local energy does not
change. Thus the energy change is generally expressed as
\begin{eqnarray}
E_{i,j}^{t+1/2} - E_{i,j}^{t} 
&=& -2\sigma_{i,j}^{t}(\sigma_{i-1,j}^{t}+\sigma_{i,j-1}^{t} )
      \,\, \delta(\sigma_{i,j}^{t+1/2} + \sigma_{i,j}^{t})  
\nonumber \\
&=& \sigma_{i-1,j}^{t} ( \sigma_{i,j}^{t+1/2}-\sigma_{i,j}^{t})
   +\sigma_{i,j-1}^{t} ( \sigma_{i,j}^{t+1/2}-\sigma_{i,j}^{t}) ,
\label{tplushalf}
\end{eqnarray}
where $\delta(x)$ is Kronecker's delta
\begin{equation}
  \label{DELTA}
  \delta(x)=\left\{\begin{array}{ll}
1 & \quad {\rm if}\quad x=0 \\
0 & \quad {\rm otherwise}
\end{array}\right. 
\label{CR}
\end{equation}
and we have used the equality $\delta(x+y)=(1-xy)/2$ that holds for
$x,y\in\{+1,-1\}$. 
The energy difference between $t+1/2$ and $t+1$ is calculated in the
same manner, and we obtain
\begin{eqnarray}
E_{i,j}^{t+1} - E_{i,j}^{t+1/2} 
&=&   2 \sigma_{i,j}^{t+1/2}\sigma_{i+1,j}^{t+1/2} \,\,
      \delta(\sigma_{i+1,j}^{t+1} + \sigma_{i+1,j}^{t+1/2}) 
 +   2 \sigma_{i,j}^{t+1/2}\sigma_{i,j+1}^{t+1/2} \,\,
      \delta(\sigma_{i,j+1}^{t+1} + \sigma_{i,j+1}^{t+1/2}) \nonumber \\
&=& \sigma_{i,j}^{t+1/2}(\sigma_{i+1,j}^{t+1/2} - \sigma_{i+1,j}^{t+1} )
 +  \sigma_{i,j}^{t+1/2} (\sigma_{i,j+1}^{t+1/2} - \sigma_{i,j+1}^{t+1} ) ,
\label{tplusone}
\end{eqnarray}
Combining Eqs.\ (\ref{tplushalf}) and (\ref{tplusone}), 
we obtain the following expression for the energy difference between
$t$ and $t+1$.
\begin{eqnarray}
E_{i,j}^{t+1} - E_{i,j}^{t} 
&=& \sigma_{i-1,j}^{t}( \sigma_{i,j}^{t+1}-\sigma_{i,j}^{t})
 +  \sigma_{i,j}^{t+1}( \sigma_{i+1,j}^{t}-\sigma_{i+1,j}^{t+1}) \nonumber \\
& & \mbox{}+\sigma_{i,j-1}^{t}( \sigma_{i,j}^{t+1}-\sigma_{i,j}^{t})
 +  \sigma_{i,j}^{t+1}( \sigma_{i,j+1}^{t}-\sigma_{i,j+1}^{t+1}), 
\label{DIFF}
\end{eqnarray}
where we have used the fact that $\sigma_{i,j}^{t+1/2}=\sigma_{i,j}^{t}$ for
odd $(i,j)$ and $\sigma_{i,j}^{t+1/2}=\sigma_{i,j}^{t+1}$ for even $(i,j)$.
Because the total energy is conserved,
Eq.\  (\ref{DIFF}) must represent the equation of continuity, 
\begin{eqnarray}
\label{CONT}
E_{i,j}^{t+1} - E_{i,j}^{t} &=& -J_{i+1,j,x}^t+J_{i,j,x}^t
                                -J_{i,j+1,y}^t+J_{i,j,y}^t ,
\end{eqnarray}
where $J_{i,j,\alpha}^t$ ($\alpha=x$ or $y$) denotes the $\alpha$
component of the energy flux at site $(i,j)$ at time $t$. 
Comparing Eqs.\ (\ref{DIFF}) and (\ref{CONT}), 
we find that the components of the energy flux are given as
\begin{eqnarray}                        
J_{i,j,x}^{t} &=& 
\sigma_{i-1,j}^{t}(\sigma_{i,j}^{t+1}-\sigma_{i,j}^{t}),
\nonumber  \\
J_{i+1,j,x}^{t} &=& 
\sigma_{i,j}^{t+1} (\sigma_{i+1,j}^{t+1}-\sigma_{i+1,j}^{t}) ,
\nonumber  \\
J_{i,j,y}^{t} &=& 
\sigma_{i,j-1}^{t}(\sigma_{i,j}^{t+1}-\sigma_{i,j}^{t}),
\nonumber  \\
J_{i,j+1,y}^{t} &=& 
\sigma_{i,j}^{t+1} (\sigma_{i,j+1}^{t+1}-\sigma_{i,j+1}^{t}).  
\nonumber 
\end{eqnarray}
The same argument can also be applied to the case where site $(i,j)$ 
is odd.
As the result we arrive at the following expressions for
the energy flux.  If site $(i,j)$ is even,
\begin{eqnarray}
J_{i,j,x}^t &=& 
\sigma_{i-1,j}^{t}(\sigma_{i,j}^{t+1}-\sigma_{i,j}^{t}), \label{EVJ}\\
J_{i,j,y}^t &=&
\sigma_{i,j-1}^{t}(\sigma_{i,j}^{t+1}-\sigma_{i,j}^{t}) ,
\label{FLE}           
\end{eqnarray}
and if site $(i,j)$ is odd,
\begin{eqnarray}
J_{i,j,x}^t &=& 
\sigma_{i-1,j}^{t+1}(\sigma_{i,j}^{t+1}-\sigma_{i,j}^{t}), \label{ODJ}\\
J_{i,j,y}^t &=&
\sigma_{i,j-1}^{t+1}(\sigma_{i,j}^{t+1}-\sigma_{i,j}^{t}).
\end{eqnarray}

\section{Thermal Conduction under a Gradient of the Temperature}

In this section, we report numerical results on energy transport in 
the CCA obtained by a direct simulation.
We took the systems of size $L\times L$ where $L$ varies 
from 10 to 300.
The periodic boundary condition was imposed on the $y$ direction.  
At the ends in the $x$ direction, two heat reservoirs,
one at temperature $T_{\rm L}$ and the other at $T_{\rm R}$, 
were attached as shown in Fig.\ 1.  
Each heat reservoir consisted of spins on two vertical lines,
where the spins on a sublattice 
were simultaneously updated by the use of
the Monte-Carlo method with the heat-bath algorithm.
We have numerically confirmed that if the two
heat reservoirs have an identical temperature the system
relaxes to the equilibrium state at that temperature.  
This relaxation to the equilibrium was also observed in the case 
where only one reservoir was attached to the system.    

Energy transport occurs when the left and right reservoirs
have different temperatures.  
It is found that the relaxation time to a stationary state is
very long at low temperatures 
below $T_{\rm C}=2/\log(1+\sqrt{2})\simeq 2.270$, 
while it is rather short at high temperatures.

The following two cases are examined with particular care.
One is the case where both the temperatures 
$T_{\rm L}$ and $T_{\rm R}$ are higher than $T_{\rm C}$, 
$T_{\rm L}=5.0$ and $T_{\rm R}=5.5$. This is called case A. 
The other is the case of $T_{\rm L}= 2.1$ and $T_{\rm R}=2.2$, 
where both the temperatures 
are lower than $T_{\rm C}$. We call this case B. In each case,
within $10^7$ time steps 
the system of any size ($L\leq 300$) reached a stationary state where
a uniform flux in the $x$ direction is realized.
After the system reached the stationary state, 
we continued the simulation by $10^7$ more steps 
for which we took time averages of physical quantities.

First we consider the distribution of a local kinetic energy, 
$P_{i,j}(\tilde{\sigma})$.  Because the system is translation 
invariant in the vertical direction, we computed the average of 
$P_{i,j}$ over the vertical line and found that it is given by
a canonical distribution
\begin{equation}
\frac{1}{L}\sum_{j=1}^LP_{i,j}(\tilde{\sigma})\propto 
\exp(-4\beta_{i}\tilde{\sigma}),
\label{cano}
\end{equation}
where $\beta_{i}$ is a fitting parameter which is regarded as
the local inverse temperature at line $i$. 
Figures 2(a) and (b) show the distributions for case A and case B,
respectively. They clearly demonstrate the property (\ref{cano}).
Thus local equilibrium is realized and the local temperatures 
are well defined.

Let $T_i$ denote the temperature at horizontal position $i$, namely
$T_{i}=\beta_{i}^{-1}$.
We plotted $T_i$ as a function of $x=i/L$ for various $L$s in
Fig.\ 3(a) and (b), which correspond to cases A and B, respectively.  
Clearly the scaling limit
\begin{equation}
  T(x)=\lim_{L\rightarrow\infty}T_{[Lx]} ,
\label{scaling}
\end{equation}
where $[Lx]$ means the integer part of $Lx$, exists 
and is smooth in both the cases A and B.  

Next we observed the total energy flux per row in the stationary state
\begin{eqnarray}
\frac{\overline{J_{{\rm tot},x}}}{L}   
&=& \frac{1}{L}\sum_{i,j=1}^{L} \overline{J_{i,j,x}},
\end{eqnarray}
where $J_{{\rm tot},x}$ is the total energy flux in the $x$ direction and the
bars mean the time average in the stationary state.
If the Fourier heat law is realized, this quantity must converge to 
a nonzero constant in the limit $L\rightarrow\infty$ with 
$T_{\rm L}$ and $T_{\rm R}$ fixed,
because then this quantity is written as
\begin{equation}
\frac{\overline{J_{{\rm tot},x}}}{L} 
= -\int_{T_{\rm L}}^{T_{\rm R}} \kappa (T) dT
\end{equation}
with use of the thermal conductivity $\kappa(T)$.
We utilized this property to judge whether the Fourier heat law is satisfied
or not.

In Fig.\  4, the $L$ dependence of $J_{{\rm tot},x}/L$ is
shown for various temperature values.
The figure shows that the size dependence disappears
in the large systems.
Thus we conclude that the Fourier heat law is realized in a wide
range of temperatures including the critical point.

Moreover, $J_{{\rm tot},x}/L$ has a finite value
and changes smoothly around the critical temperature. 
This observation suggests that the thermal conductivity has no strong
singularity at $T_{\rm C}$. 
However, we can treat not the thermal conductivity itself but the 
integration of it in the present method and a possible singularity, 
if any, is hardly observed.
Thus in the next section we investigate the thermal conductivity in 
the bulk at a given temperature with use of the Kubo formula.

\section{Thermal Conductivity computed via the Kubo Formula}

According to the Kubo formula, the thermal conductivity is equal to the
summation of the equilibrium autocorrelation functions of the energy flux as 
\begin{equation}
 \kappa(T) = \frac{1}{NT^2}\sum_{t=0}^{\infty} 
\langle J_{{\rm tot}, x}^0J_{{\rm tot},x}^t\rangle 
\left(1-\frac{1}{2}\delta_{t,0}\right),
\label{KF} 
\end{equation}
where $J_{{\rm tot},x}^t=\sum_{i,j}J_{i,j,x}^t$ is the total energy flux 
in the $x$ direction at time $t$, $\langle\cdots\rangle$ means the 
equilibrium ensemble average at temperature $T$, 
and $N$ is the total number of sites.
This formula is proved for the CCA in Appendices A and B. 

We numerically computed the autocorrelation functions
$\langle J_{{\rm tot}, x}^0J_{{\rm tot},x}^t\rangle$
for $t\leq 150$ in the CCA under the periodic boundary conditions in the
$x$ and $y$ directions.  Initial conditions were randomly generated by a
Monte-Carlo method with temperature $T$.
We denote the partial Kubo sum up to time $t$ by $\kappa^t$, namely
\begin{equation}
 \kappa^t = \frac{1}{NT^2}\sum_{t'=0}^{t}  
\langle J_{{\rm tot}, x}^0J_{{\rm tot},x}^{t'}\rangle 
\left(1-\frac{1}{2}\delta_{t',0}\right).
\end{equation}
Figure 5 shows numerically computed $\kappa^t$ in the system of 
size $200\times 200$ at various temperatures.
It is observed that the summation
converges by $t=30$ for every temperature.
At temperatures above $T_{\rm C}$, the sum monotonically increases 
and tends to a constant exponentially fast. At low temperatures the
monotonicity is lost and significant fluctuations appear.  However,
$\kappa^t$ still reaches a convergence by $t=10$.

Figure 6 shows the thermal conductivity thus obtained and that computed via
the direct measurement of energy flux as explained in the previous section.  
Both the results agree with each other very well.
From the figure we know that the
thermal conductivity has a peak at $T\sim 2.70$, which is slightly above the
critical temperature $T_{\rm C}$.  Above the peak value, the thermal
conductivity gradually decreases and tends to zero in the high temperature
limit.  Below the peak value, the conductivity shows a remarkable change 
around $T_{\rm C}$ and reaches nearly zero at $T=2.0$. 
Detailed measurements were done near the critical temperature $T_{\rm C}$ and
the results are shown in Fig.\ 7. This figure shows that $\kappa(T)$ appears
continuous and smooth at the critical point, 
though the magnitude of the change is large.  
Because little size dependence is seen when $L\geq 100$, we can
conclude that at least no divergence or no vanishment of $\kappa(T)$
occurs at $T_{\rm C}$.   
Of course we cannot deny the possibility of singularity or discontinuity in
a higher derivative. 

\section{Mean-Field Analysis of Thermal Conductivity}

In this section, we estimate the equal time correlation function of
the heat flow using a mean-field approximation and discuss  
its temperature dependence. 
This quantity is the first term of the
Kubo formula Eq.\ (\ref{KF}), namely $\kappa^{0} (T)$, 
and thus we can obtain some information on the temperature dependence of
the thermal conductivity.

As derived in Appendix B, $\kappa^0(T)$ is expressed in terms of 
an average of the total flow $J_{{\rm tot},x}$ in the local equilibrium as
\begin{equation}
\kappa^0 (T) 
    = \frac{1}{2 T^2 N} \langle J_{{\rm tot},x} J_{{\rm tot},x} \rangle
    = \lim_{|T_{\rm L}-T_{\rm R}|\to 0} 
 \frac{ \langle J_{{\rm tot},x}\rangle_{\rm le}}{N(T_{\rm L}-T_{\rm R})},  
\label{RHOTOT} 
\end{equation}
where $\langle\cdots\rangle_{\rm le}$ denotes the average with
respect to the local equilibrium product measure (\ref{LEQ}) 
with the right reservoir temperature $T_{\rm R}$ and 
the left reservoir temperature $T_{\rm L}$.

First we consider the quantity $\langle J_{i,j,x} \rangle_{\rm le}$ at
an even site $(i,j)$.  Denoting the local equilibrium measure by
$\rho_{\rm le}$, we have
\begin{equation}
\langle J_{i,j,x} \rangle_{\rm le}= 
\sum_{\{\sigma,\tilde{\sigma} \}} J_{i,j,x}\rho_{\rm le}. 
\end{equation}
Substituting Eq.\ (\ref{EVJ}) into $J_{i,j,x}$, we can express this as
\begin{eqnarray}
\langle J_{i,j,x}  \rangle_{\rm le} 
&=& 
\sum_{\{\sigma,\tilde{\sigma}\}}
(\sigma_{i-1,j} \sigma'_{i,j} 
-\sigma_{i-1,j}\sigma_{i,j}  )\rho_{\rm le}   \nonumber \\
&=& -2 {\sum_{\{\sigma,\tilde{\sigma} \}}}^*
\sigma_{i-1,j} \, \sigma_{i,j} \, \rho_{\rm le},
\label{sumstar}
\end{eqnarray}
where $\sigma'_{i,j}$ denotes the updated spin value at the even site 
$(i,j)$ and ${\sum}^{*}$ means the summation over the configurations 
in which the spin $\sigma_{i,j}$ can flip.
Whether the spin flip occurs or not depends on 
the spin and the momentum variable at $(i,j)$ and
the sum of the spin values on the nearest neighbor sites,
\begin{equation}
  h=\sigma_{i-1,j} + \sigma_{i+1,j} + \sigma_{i,j-1} + \sigma_{i,j+1}.
\end{equation}
Specifically, the spin flip is possible in the following configurations: 
\begin{eqnarray}
(h, \sigma_{i,j}, \tilde{ \sigma }_{i,j} ) 
 =  \left\{ 
          \begin{array}{l}
       ( \pm 4, \pm  1, (3 ,2))     \\
       ( \pm 2, \pm  1, (3 ,2, 1))  \\
       ( \pm 0, \pm  1, (3 ,2, 1, 0))  \\
       ( \pm 4, \mp  1, (1, 0))     \\
       ( \pm 2, \mp  1, (2, 1, 0)) \,\, . \\
             \end{array}   
              \right. 
\end{eqnarray}

Because the summation (\ref{sumstar}) must be taken over the 
configurations for the whole system,
it is difficult to be carried out exactly.
Thus we consider the following mean field approximation.
In this approximation the spin variables at the next nearest neighbor 
sites are replaced by their average values.  Those average values should
depend only on the horizontal position and not on the vertical position,
since 
the local equilibrium measure is translation invariant in the $y$ direction.
Thus the average concerning the local equilibrium measure is replaced by the
average concerning the following measure
\begin{equation}
\frac{ P(\sigma,\tilde{\sigma},\langle\sigma\rangle)}
{\sum_{\sigma,\tilde{\sigma}}P(\sigma,\tilde{\sigma},\langle\sigma\rangle)},
\label{mfmeas}
\end{equation} 
where
\begin{eqnarray} 
P (\sigma , \tilde{\sigma}, \langle\sigma\rangle) 
&=& \exp(-4\beta_i \tilde{\sigma}_{i,j})
\exp[a\sigma_{i-1,j}+b\sigma_{i,j-1}+c\sigma_{i,j+1}+d\sigma_{i+1,j}   
\nonumber \\
& & \mbox{}+\sigma_{i,j}(\beta_{i-1}'\sigma_{i-1,j}+\beta_i\sigma_{i,j+1} 
    +\beta_i\sigma_{i,j-1}+\beta_i\sigma_{i+1,j})]
\end{eqnarray}
with
\begin{eqnarray} 
a &=& \beta_{i-2} \langle\sigma \rangle_{i-2} 
+\beta_{i-1} \langle\sigma \rangle_{i-1} +\beta_{i-1} 
\langle\sigma \rangle_{i-1} , \nonumber \\
b &=& \beta_{i-1} \langle\sigma \rangle_{i-1} +\beta_{i} 
\langle\sigma \rangle_{i} +\beta_{i} 
\langle\sigma \rangle_{i+1} , \nonumber \\
c &=& \beta_{i-1} \langle\sigma \rangle_{i-1} +\beta_{i} 
\langle\sigma \rangle_{i} +\beta_{i} 
\langle\sigma \rangle_{i+1} , \nonumber \\
d &=& \beta_{i+1} \langle\sigma \rangle_{i+1} +\beta_{i+1} 
\langle\sigma \rangle_{i+1} +\beta_{i+1} 
\langle\sigma \rangle_{i+2}. \nonumber
\end{eqnarray}
The summation in the denominator of (\ref{mfmeas}) is taken over possible
values of
$\sigma_{i,j}$, $\tilde{\sigma}_{i,j}$ and the nearest neighbor spins.
Here $\beta_{i}$ is the inverse temperature at horizontal position $i$ and 
$\beta_{i-1}'$ takes the same value as $\beta_{i-1}$.
We introduced $\beta_{i-1}'$ for later convenience.
$\langle\sigma \rangle_{i}$ denotes the local equilibrium 
value of the spin variable at horizontal position $i$.

Under the above approximation, $\langle J_{i,j,x}\rangle_{\rm le}$ is 
represented as 
\begin{equation}
\langle J_{i,j,x}\rangle_{\rm le}
\simeq -\frac{2}{Z}\frac{\partial Z^{*}}{\partial \beta_{i-1}'} ,
\end{equation}
where $Z$ and $Z^*$ are defined by
\begin{eqnarray}
Z^{*} &=& \sum_{\sigma ,\tilde{\sigma } }^{} {}^* 
P (\sigma , \tilde{\sigma}, \langle\sigma\rangle  ) ,
\nonumber \\
Z     &=& \sum_{\sigma , \tilde{\sigma }} 
P (\sigma , \tilde{\sigma}, \langle\sigma\rangle  ).\nonumber 
\end{eqnarray}
With straightforward calculation $Z^{*}$ is obtained as,
\begin{eqnarray}
Z^{*} 
&=& 
2 e^{\beta_{i-1}'-\beta_{i}} \, ( e^{-4\beta_{i}} + e^{-8 \beta_{i}}) 
\cosh(a+b+c+d)   \nonumber \\
&& \mbox{}+2( e^{-10 \beta_{i} } + e^{-6 \beta_{i} } + e^{-2 \beta_{i} } )
\left\{ e^{ \beta_{i}    - \beta_{i-1}' }\cosh(a-b-c-d) 
 + e^{ \beta_{i-1}' - \beta_{i}}\cosh(a-b+c+d)  \right.
\nonumber \\
&&  \left.
          + e^{ \beta_{i-1}' - \beta_{i} }  {\rm cosh }(a+b-c+d)
          + e^{ \beta_{i-1}' - \beta_{i} }  {\rm cosh }(a+b+c-d)   \right\} 
\nonumber \\
&& \mbox{}+2(e^{-12\beta_{i}} +e^{-8\beta_{i}}+e^{-4\beta_{i}}+1)  \nonumber \\
&&\times
\left\{ 
e^{\beta_{i-1}'-\beta_{i}}\cosh(a+b-c-d)+
e^{\beta_{i-1}'-\beta_{i}}\cosh(a-b-c+d)
+e^{\beta_{i-1}'-\beta_{i}}\cosh(a-b+c-d)   \right. \nonumber \\
& &\mbox{}+\left.
e^{ - \beta_{i}' - \beta_{i} }\cosh(-a-b+c+d) 
+ e^{ - \beta_{i}' - \beta_{i}}\cosh(-a+b-c+d) 
+ e^{ - \beta_{i}' - \beta_{i} }\cosh(-a+b+c-d)   
\right\} \nonumber \\
&&\mbox{}+ 2 e^{-\beta_{i-1}'+\beta_{i}} (e^{-4\beta_{i}}+e^{-8\beta_{i}}) 
\cosh(a+b+c+d) \nonumber \\
&&\mbox{}+ 2( e^{-10 \beta_{i} } + e^{ -6 \beta_{i} }+ e^{ -2 \beta_{i} } )
\left\{ 
  e^{ \beta_{i-1}' - \beta_{i} }   \cosh(-a+b+c+d) 
+ e^{ - \beta'_{i-1} + \beta_{i} } \cosh(a-b+c+d)  \right. \nonumber \\
&& \left.
+ e^{ - \beta'_{i-1} + \beta_{i} } \cosh(a+b-c+d) 
+ e^{ - \beta'_{i-1} + \beta_{i} } \cosh(a+b+c-d)  \right\} .
\end{eqnarray} 
In the first order of $ \Delta T ( := T_{i} - T_{i-1} )$, 
$\frac{\partial Z^{*}}{\partial \beta'_{i-1}} 
\left|_{  \beta'_{i-1}=\beta_{i}+ \beta_i^2 \Delta T  } \right. $ 
can be simplified as
\begin{eqnarray}
\left. \frac{\partial Z^{*}}{\partial \beta'_{i-1}} 
\right|_{\beta'_{i-1}= \beta_{i}+ \beta_i^2 \Delta T} 
&\simeq & \frac{4 \Delta T}{T^{2}}
\{ (e^{-4\beta_{i}}+e^{-8\beta_{i}})\cosh(12\beta_{i} 
\langle\sigma\rangle_{i}  )  
\nonumber \\
&& \mbox{}+4( e^{-10\beta_{i}} + e^{-6\beta_{i}} + e^{-2\beta_{i}} )
\cosh(6\beta_{i}\langle\sigma\rangle_{i} ) 
+ 3(1+  e^{-8\beta_{i}})(1 + e^{-4\beta_{i}})
\}. 
\end{eqnarray} 
$Z$ is also calculated as 
\begin{equation}
Z 
= 2^{4} \left(  \cosh^{4}(3 \beta_{i}\langle\sigma\rangle_{i} 
+\beta_{i})
    + \cosh^{4}(3\beta_{i}\langle\sigma\rangle_{i} 
-\beta_{i}) 
            \right)
(1+e^{-4\beta_{i}})(1 + e^{-8\beta_{i}}).
\end{equation} 
Thus we arrive at the approximate formula for $\kappa^0(T)$,
\begin{eqnarray}
\kappa^{0} (T)
&=&\lim_{\Delta T\to 0}
\frac{ \left\langle  J_{{\rm tot},x} /N \right\rangle_{\rm le}}{\Delta T} 
\nonumber \\
&\simeq & 
\frac{ \cosh(12\beta\langle\sigma\rangle )
       \cosh( 2\beta\langle\sigma\rangle )
+ 2(1+2\cosh 4\beta  )\cosh (6\beta\langle\sigma\rangle )
+ 6\cosh 2\beta \cosh 4\beta }
{ 4 T^{2} \left[  \cosh^{4}(3\beta\langle\sigma\rangle +\beta )
+      \cosh^{4}(3\beta\langle\sigma\rangle -\beta )  \right]
\cosh 2\beta \cosh 4\beta  }, \nonumber \\
\label{kappazero} 
\end{eqnarray} 
where in the limit of $\Delta T \rightarrow 0$ the system becomes uniform
and we identify $\langle J_{{\rm tot},x}/N\rangle_{\rm le}$ with 
$\langle J_{i,j,x}\rangle_{\rm le}$ and 
put $\langle\sigma\rangle := \langle\sigma\rangle_{i}$.
In the high temperature limit, using $\langle\sigma\rangle =0 $, we have
\begin{eqnarray}
\kappa^{0}(T) &\sim & \frac{13}{8}\frac{1}{T^2},
\end{eqnarray}
while in the low temperature limit, using $\langle\sigma\rangle =1 $,
\begin{eqnarray}
\kappa^{0}(T) &\sim & \frac{4}{T^2} e^{-8/T }.
\end{eqnarray}
These asymptotic forms are the same as obtained by Harris and 
Grant\cite{HG88}.

However, the formula (\ref{kappazero}) gives more information about 
overall temperature dependence.
Although the present approximation is not good near the critical point,
within this approximation we find that $\kappa^{0}(T)$ is continuous
but shows a cusp at the mean-field critical temperature 
$T_{\rm M}\simeq 3.5$ because $\langle\sigma\rangle
\propto ( T_{\rm M} -T )^{\frac{1}{2}}$.
In Fig.\ 8 we compare the mean-field results with the numerical ones obtained 
in the previous section.  In the high temperature region both the results
agree with each other, while discrepancies appear at low temperatures. 
This is partly due to the difference between the mean-filed critical
temperature and the true critical temperature $T_{\rm C}\simeq 2.27$.
In addition the simulation results have no cusp and 
actually change smoothly.

In Fig.\ 9 we show $\kappa(T)$ and
$3.5\times \kappa^0(T)$ both of which are numerically obtained 
from the equilibrium autocorrelation functions of the energy flux.
This figure shows that $\kappa^{0}(T)$ is nearly proportional to 
$\kappa (T)$ in high temperatures.  This implies that the autocorrelation
functions of the energy flux are similar in this temperature region, 
which is also perceived by comparing the two curves for $T=3.0$ and $T=3.5$ 
in Fig.\ 5.

\section{Heat Conduction in the Q2R} 

As a simplified variant of the CCA, the Q2R was devised and some equilibrium
and dynamical features were investigated\cite{V84,H86,CH87}.  There
are no momentum variables in the Q2R, where a spin flips only when the sum 
of the nearest-neighbor spins is zero.  Despite the simplicity of the model,
it is known that the critical behavior for the magnetization can be simulated
by this model. 

We have performed direct simulations of the Q2R in contact with two heat
reservoirs at different temperatures in almost the same manner as 
in Sec.\ 3.
Heat reservoirs were realized by the same algorithm as shown in Fig.\ 1.
The temperatures of the reservoirs were set as $T_{\rm L}=6.0$ 
and $T_{\rm R}=10.0$.  Here we took
quasi one-dimensional systems of size $L\times 10$ with various $L$s. 
Simulation time for each size is $5\times 10^7$ time steps.
The expressions for the energy flux (\ref{EVJ}) and (\ref{ODJ}) can be 
used without changes because they do not contain momentum variables. 
For the same reason, in the Q2R we cannot determine local temperature 
from the distribution of local kinetic energy as was done in the CCA.  
Thus we plotted local energies in the stationary state for various
system sizes in Fig.\ 10.  As in the Creutz model, the Q2R also shows a
smooth energy profile in the scaling limit (\ref{scaling}).  System-size
dependence of the total energy flux is shown in Fig.\ 11.  The total energy
flux converges to a nonzero finite value in the limit $L\rightarrow\infty$
and it demonstrates that the Q2R has a normal thermal conductivity at least
when the temperatures are sufficiently high.  This means that the normal
thermal conductivity in the CCA is not caused by the presence of the momentum
terms. 

The thermal conductivity was carefully calculated with use of the Kubo
formula in a system of size $100\times 100$ at temperatures around 
$T_{\rm C}$.  The result is shown in Fig.\ 12, which exhibits
similar behavior to the CCA.  The thermal conductivity
shows a remarkable change near $T_{\rm C}$ but seems continuous and smooth.
This result disagrees with Costa and Herrmann\cite{CH87}, where they reported
that energy flux vanished at the critical point and no transport occurred 
below the critical point.  This discrepancy may be attributed to the
differences in system sizes and heat reservoirs in their and our systems.
In \cite{CH87}, the distance between the reservoirs is 20, which may be too
small to obtain bulk thermal conductivity.  Their heat reservoir is
deterministic and keeps energy a constant in the boundary layer representing
the reservoir.  Thus the motion of the total system must eventually 
turn into periodic.  Because energy flow rarely occurs in low temperature,
such simple dynamics may not be able to generate it, whereas our reservoirs
are stochastic and rare events can happen.  Another possible interpretation
is that they misunderstand the great change of thermal conductivity around
the critical point as vanishing. 

In addition, Costa and Herrmann reported two different types of transport
processes. One is normal diffusion and the other is a systematic transport
called ``highway.''  The latter causes a ballistic transport.  However, we
did not find such ballistic transport in our simulations.  This is also
attributed to the differences in heat reservoirs and system sizes.
The highway is characteristic of their deterministic reservoirs and moreover
the fraction of highways decreases to zero as the system size increases.

At the end of this section, we mention the one-dimensional Q2R dynamics.
If $i$ is even, 
the spin value of site $i$ at time $t+1$ is expressed in terms of spin 
variables at time $t$ as 
\begin{eqnarray}
\sigma_{i}^{t+1} = \sigma_{i}^{t+1/2} 
&=& \sigma_{i}^{t} - 2 \sigma_{i}^{t} \,
\delta \left( \sigma_{i-1}^{t} + \sigma_{i+1}^{t}  \right) \nonumber \\
&=& \sigma_{i-1}^{t}  \sigma_{i}^{t}  \sigma_{i+1}^{t} 
\label{S2I},  
\end{eqnarray}
In the same manner, if $i$ is odd
\begin{eqnarray}
\sigma_{i}^{t+1} &=& \sigma_{i}^{t+1/2} 
- 2 \sigma_{i}^{t+1/2} \,
\delta \left( \sigma_{i-1}^{t+1/2} + \sigma_{i+1}^{t+1/2}  \right) 
\nonumber \\
&=& \sigma_{i-2}^{t}  \sigma_{i-1}^{t}  \sigma_{i}^{t} 
    \sigma_{i+1}^{t}  \sigma_{i+2}^{t} .
\label{S2I1}
\end{eqnarray}
Defining local energy of site $i$ at time $t$ as
\begin{eqnarray}
E_{i}^{t} &=& -\sigma_{i}^{t} \sigma_{i+1}^{t} ,
\end{eqnarray}
we obtain the following relation for the local energy using 
Eqs.(\ref{S2I}) and (\ref{S2I1}). Namely if $i$ is even,
\begin{equation}
E_{i}^{t+1} = E_{i+2}^{t} ,
\end{equation}
and if $i$ is odd,
\begin{eqnarray}
E_{i}^{t+1} &=& E_{i-2}^{t} .
\end{eqnarray}
Therefore, the energy transport in one-dimensional Q2R is
ballistic and the Fourier heat law is not satisfied.
Thus we have found that the dimensionality has an important role for
the Fourier heat law in the Q2R.

\section{Summary and Discussion}
In this paper we have studied the thermal conduction in the CCA 
with two methods. One is the direct measurement of the heat flux 
under a temperature gradient. 
The other is the use of the Kubo formula. 
The former revealed that the assumption of local equilibrium is 
satisfied and that Fourier heat law is realized 
in a wide range of temperatures.  The thermal conductivity was 
carefully calculated near the critical point by the latter
method and the results show 
no singularity for $\kappa(T)$ at $T_{\rm C}$. 

How a thermal conductivity behaves at $T_{\rm C}$ is a highly nontrivial
problem.  Harris and Grant \cite{HG88} and Costa and Herrmann \cite{CH87} 
both argued that the thermal conductivity vanishes at the critical point.  
On the other hand, the autocorrelation of the total energy flux might show a
slow decay due to the critical slowing down.  Then the thermal conductivity
might be divergent at $T_{\rm C}$.  Our present result shows either is not
the case.

The present result does not mean that there is no singularity 
in energy transport at $T_{\rm C}$.
The Fourier heat law means that the macroscopic motion of energy
density obeys the diffusion equation with diffusion constant 
$D(T)=\kappa(T)/C(T)$, where $C(T)$ is the specific heat.  
The present result shows that $\kappa(T_{\rm C})$ is finite 
while $C(T)$ diverges to infinity at $T_{\rm C}$.  
Thus the diffusion constant $D(T)$ vanishes at $T_{\rm C}$.

We evaluated the equal time correlation of the heat flow by the use of
mean-field approximation.  This quantity is the first term in the Kubo 
formula and we can obtain a rough estimate for $\kappa(T)$.
In the high and low temperature limits, our approximation reproduces 
the result by Harris and Grant\cite{HG88}.

Similar calculations were also done for the Q2R, a simplified variant 
of the CCA.  The results obtained are almost the same as in the CCA. 
The normal thermal conductivity was found and it was continuous and 
smooth at the critical point.  This proves that the existence of the 
momentum terms is not relevant to the normal thermal conductivity.
On the other hand, the dimensionality is important.  Energy transport is
ballistic in the one-dimensional Q2R.  Such importance of the dimensionality
was reported also in \cite{Lepri}.

The similarity of the thermal conductivities in the CCA and the Q2R also 
implies that the smooth change of $\kappa$ at the critical point is rather 
generic.  To investigate to what extent this behavior is generic, however,
we must examine other dynamical systems with a critical point.
It is a future problem.   

\section*{Acknowledgment}
We gratefully acknowledge partial financial 
support from Grant-in-Aid for Scientific
Research from the Ministry of Education, Science and Culture.
Numerical computation in this work was carried out at the Yukawa Institute
Computer Facility.

\appendix
\section{}

In this and the next Appendices, we derive the Kubo formula
(\ref{KF}) for the CCA.  We denote a state of the total system
by $\omega=(\omega_{i,j})$, where
$\omega_{i,j}=(\sigma_{i,j},\tilde{\sigma}_{i,j})$, and 
the transformation  
from the state at time $t$, $\omega^t$, 
to that at time $t+1$, $\omega^{t=1}$,  by $\Omega$ as
\begin{equation}
  \omega^{t+1}=\Omega(\omega^t)
\end{equation}
Then, the time evolution of any function $F(\omega)$ is represented by
\begin{equation}
  F^{t+1}(\omega)=F^{t}(\Omega(\omega))
\end{equation}
and $F^0(\omega)=F(\omega)$, and the time evolution of a measure
$\rho(\omega)$ by
\begin{equation}
  \rho^{t+1}(\omega)=
\sum_{\omega'}\delta(\omega,\Omega(\omega'))\rho^t(\omega')
\end{equation}
and $\rho^0(\omega)=\rho(\omega)$, where 
\begin{equation}
  \delta(\omega,\omega')=\left\{
\begin{array}{ll}
1 & \quad {\rm ~if~}\omega=\omega' \\
0 & \quad {\rm ~if~}\omega\neq\omega'
\end{array}
\right.
\end{equation}

Now we define the total flux $J_{{\rm tot},x}(\omega)$ by
\begin{equation}
  J_{{\rm tot},x}(\omega)=\sum_{i,j}J_{i,j,x}(\omega)
\end{equation}
where $J_{i,j,\alpha}(\omega)$ ($\alpha=x$ or $y$)is 
the $\alpha$ component of the energy flux 
at site $(i,j)$ when the system is in state $\omega$.
We assume that the initial measure $\rho^0$ equals 
the local equilibrium measure $\rho_{\rm le}$ defined by
\begin{equation}
  \label{LEQ}
  \rho_{\rm le}(\omega)=\frac{1}{Z_{\rm le}}
\prod_{i,j}e^{-\beta_iE_{i,j}(\omega)},
\end{equation}
where $E_{i,j}(\omega)$ is the local energy around site $(i,j)$ in
state $\omega$.  $Z_{\rm le}$ denotes the normalization constant
\begin{equation}
  Z_{\rm le}=\sum_{\omega}\prod_{i,j}e^{-\beta_iE_{i,j}(\omega)}.
\end{equation}
The parameter $\beta_i$ is the local inverse temperature at the $i$th
column.  We consider the temperature variation in the $x$ direction 
only.  If the temperature is uniform and all the $\beta_i$s equal 
a value $\beta$, $\rho_{\rm le}$ becomes the equilibrium measure of
temperature $T=\beta^{-1}$.
The average of function $F(\omega)$ with respect to the
local equilibrium measure is written as
\begin{equation}
  \label{LEQAVE}
\langle F\rangle_{\rm le}=\sum_{\omega}F(\omega)\rho_{\rm le}(\omega)
\end{equation}
Similarly we write the equilibrium average as 
$\langle F\rangle_{\rm  eq}$.

In the following, we calculate the local equilibrium average of the
total flux at time $t$
\begin{eqnarray}
\langle J_{{\rm tot},x}^t\rangle_{\rm le}&=&
\langle J_{{\rm tot},x}^0\rangle_{\rm le}+\sum_{t'=0}^{t-1}
\langle J_{{\rm tot},x}^{t'+1}-J_{{\rm tot},x}^{t'}\rangle_{\rm le} \\
&=& \langle J_{{\rm tot},x}\rangle_{\rm le}+
\sum_{t'=0}^{t-1}\sum_{\omega}
(J_{{\rm tot},x}^{t'+1}(\omega)-J_{{\rm tot},x}^{t'}(\omega))
\rho^0(\omega) \\
&=& \langle J_{{\rm tot},x}\rangle_{\rm le}+
\sum_{t'=0}^{t-1}\sum_{\omega}
J_{{\rm tot},x}^{t'}(\omega)(\rho^1(\omega)-\rho^0(\omega))
\label{JTOT}
\end{eqnarray}
In the last equality, we have used the identity
\begin{equation}
\sum_\omega F^{t+1}(\omega)\rho^0(\omega)=
\sum_\omega\sum_{\omega'}F^{t+1}(\omega)
\delta(\omega',\Omega(\omega))\rho^0(\omega)
=\sum_{\omega'}F^{t}(\omega')\rho^1(\omega).
\end{equation}
Utilizing the equation of continuity
\begin{equation}
  E_{l,m}(\Omega(\omega))=
E_{l,m}(\omega)-J_{l+1,m,x}(\omega)+J_{l,m,x}(\omega)
-J_{l,m+1,y}(\omega)+J_{l,m,y}(\omega),
\end{equation}
$\rho^1(\omega)$ is calculated as
\begin{eqnarray}
  \rho^1(\omega)&=&\frac{1}{Z_{\rm le}}\sum_{\omega'}
\delta(\omega,\Omega(\omega'))\prod_{l,m}e^{-\beta_lE_{l,m}(\omega')}
\nonumber\\
&=&\frac{1}{Z_{\rm le}}\sum_{\omega'}
\delta(\omega,\Omega(\omega')) \nonumber \\
& &\quad\times\prod_{l,m}
\exp\left(-\beta_l[E_{l,m}(\Omega(\omega'))+J_{l+1,m,x}(\omega')
-J_{l,m,x}(\omega')+J_{l,m+1,y}(\omega')-J_{l,m,y}(\omega')]\right)
\nonumber \\
&=&
\rho_{\rm le}(\omega)\sum_{\omega'}\delta(\omega,\Omega(\omega'))
\prod_{l,m}e^{-\beta_l[J_{l+1,m,x}(\omega')-J_{l,m,x}(\omega')]}
\nonumber \\
&=&
\rho_{\rm le}(\omega)\sum_{\omega'}\delta(\omega,\Omega(\omega'))
\prod_{l,m}e^{(\beta_l-\beta_{l-1})J_{l,m,x}(\omega')}
\end{eqnarray}
Inserting the above formula into Eq.\ (\ref{JTOT}), we have
\begin{equation}
  \label{JTOT2}
\langle J_{{\rm tot},x}^t\rangle_{\rm le}=
\langle J_{{\rm tot},x}\rangle_{\rm le}
+\sum_{t'=0}^{t-1}\sum_{\omega}
J_{{\rm tot},x}^{t'}(\omega)\rho_{\rm le}(\omega)\left\{
\sum_{\omega'}\delta(\omega,\Omega(\omega'))
\prod_{l,m}e^{(\beta_l-\beta_{l-1})J_{l,m,x}(\omega')}
-1\right\}
\end{equation}
Now we formally expand the right hand side with respect 
to $\nabla T$ and obtain in $O(\nabla T)$
\begin{eqnarray}
\langle J_{{\rm tot},x}^t\rangle_{\rm le}
&\simeq&
\langle J_{{\rm tot},x}\rangle_{\rm le}
+\sum_{t'=0}^{t-1}\sum_{\omega}\sum_{\omega'}
J_{{\rm tot},x}^{t'}(\omega)\rho_{\rm eq}(\omega)
\delta(\omega,\Omega(\omega'))
\sum_{l,m}(\beta_l-\beta_{l-1})J_{l,m,x}(\omega')\nonumber \\
&\simeq&
\langle J_{{\rm tot},x}\rangle_{\rm le}
+\sum_{t'=0}^{t-1}\sum_{\omega'}
J_{{\rm tot},x}^{t'+1}(\omega')\rho_{\rm eq}(\omega')
\sum_{l,m}(\beta_l-\beta_{l-1})J_{l,m,x}(\omega')\nonumber \\
&\simeq & \langle J_{{\rm tot},x}\rangle_{\rm le}
-\frac{\nabla T}{T^2}\sum_{t'=1}^t
\langle J_{{\rm tot},x}J_{{\rm tot},x}^{t'}\rangle_{\rm eq}
\end{eqnarray}
where we have used the time invariance of the equilibrium measure,
$\rho_{\rm eq}(\Omega(\omega))=\rho_{\rm eq}(\omega)$.

As we will show in Appendix B, the following equality holds for the
first term in the right hand side
\begin{equation}
\langle J_{{\rm tot},x}\rangle_{\rm le}\simeq
- \frac{\nabla T}{2 T^{2}}\langle (J_{{\rm tot},x})^2\rangle_{\rm eq}
\label{C00}
\end{equation}
in $O(\nabla T)$.
In addition, we assume that the average energy flux goes to 
a stationary value in the limit $t\rightarrow\infty$ 
irrespective of an initial measure.
Then the stationary energy flux per site obeys the Fourier heat law
\begin{equation}
  J_{\rm st}\equiv \lim_{t\rightarrow\infty}\frac{1}{N}
\langle J_{{\rm tot},x}^t\rangle_{\rm le}=-\kappa\nabla T
\quad {\rm in~}O(\nabla T)
\end{equation}
and the thermal conductivity $\kappa$ is given by
\begin{equation}
\kappa (T) = \sum_{t=0}^{\infty }\frac{1}{ T^2 N}
\langle J_{{\rm tot},x} J_{{\rm tot},x}^{t} \rangle 
(1 - \frac{1}{2}\delta_{t,0} ),  
\end{equation}
which is the Kubo formula for the CCA.
We remark that the expansion is formal and not justified. 
The coefficient $\kappa$ might be divergent.
Currently we have no means to judge the convergence of 
the coefficient except the numerical methods.

\section{}

In this Appendix we prove the formula (\ref{C00}).
First we note that
\begin{equation}\label{JDEV}
\left.\frac{\partial}{\partial\nabla\beta}
\langle J_{{\rm tot},x}\rangle_{\rm le}\right|_{\nabla\beta=0}
=\sum_{i,j}\sum_{l,m}l
(\langle J_{i,j,x}\rangle_{\rm eq} \langle E_{l,m}\rangle_{\rm eq}
-\langle J_{i,j,x}E_{l,m}\rangle_{\rm eq}).
\end{equation}
This is obtained by a straightforward calculation.
From now on, we only deal with equilibrium averages and suffix ``eq''
will be omitted.

Since the total Hamiltonian $H$ is invariant, namely
$H(\Omega(\omega))=H(\omega)$, we have
\begin{eqnarray}
  \langle J_{i,j,x}E_{l,m}\rangle &=&
Z^{-1}\sum_{\omega}J_{i,j,x}(\omega)E_{l,m}(\omega)
e^{-\beta H(\omega)} \nonumber \\
&=&Z^{-1}\sum_{\omega}J_{i,j,x}(\omega)E_{l,m}(\omega)
e^{-\beta H(\Omega(\omega))} \nonumber \\
&=&Z^{-1}\sum_{\omega}
J_{i,j,x}(\Omega^{-1}(\omega))E_{l,m}(\Omega^{-1}(\omega))
e^{-\beta H(\omega)}
\label{JE}
\end{eqnarray}
where $\Omega^{-1}$ is the inverse operator of $\Omega$.
Denoting the operation of updating 
the even sites by $\Omega^{\rm e}$ and 
that for the odd sites by $\Omega^{\rm o}$, we can decompose 
the time evolution operator $\Omega$ as
\begin{equation}
  \Omega=\Omega^{\rm o}\circ\Omega^{\rm e}.
\end{equation}
Since we have 
$\Omega^{\rm e}\circ\Omega^{\rm e}
=\Omega^{\rm o}\circ\Omega^{\rm o}={\bf 1}$ (identity), 
the inverse operator is given as
\begin{equation}
  \Omega^{-1}=\Omega^{\rm e}\circ\Omega^{\rm o}.
\end{equation}

Let us define the shift operator $S$ by
\begin{equation}
  \label{SHIFT}
  (S\omega)_{i,j}=\omega_{i,j-1}.
\end{equation}
This means that the operator $S$ shifts the state by
one site in the $y$ direction.  
Because the shift exchanges the roles of the even and odd
sites, we have
\begin{eqnarray}
  \label{SHIFT1}
S\circ\Omega^{\rm e}&=&\Omega^{\rm o}\circ S \\
\label{SHIFT2}
S\circ\Omega^{\rm o}&=&\Omega^{\rm e}\circ S
\end{eqnarray}
Thus the inverse operator has another representation
\begin{equation}
  \Omega^{-1}=S^{-1}\circ\Omega\circ S
\end{equation}
Inserting this formula into Eq.\ (\ref{JE}) and utilizing the shift 
invariance of the Hamiltonian
(i.e., $H(S(\omega))=H(\omega)$), we can write
\begin{eqnarray}
\langle J_{i,j,x}E_{l,m}\rangle
&=&
Z^{-1}\sum_{\omega}
J_{i,j,x}(\Omega^{-1}(\omega))
E_{l,m}(S^{-1}\circ\Omega\circ S(\omega))
e^{-\beta H(\omega)} \\
&=& Z^{-1}\sum_{\omega}
J_{i,j,x}(\Omega^{-1}\circ S^{-1}(\omega))
E_{l,m}(S^{-1}\circ\Omega(\omega))e^{-\beta H(\omega)} \\
&=& Z^{-1}\sum_{\omega}
J_{i,j,x}(\Omega^{-1}\circ S^{-1}(\omega))
E_{l,m+1}(\Omega(\omega))e^{-\beta H(\omega)}
\label{JE2}
\end{eqnarray}
From the definition of the flux $J_{i,j,x}(\omega)$
\begin{equation}
 J_{i,j,x}(\omega)=\left\{
\begin{array}{ll}
\sigma_{i-1,j}(\sigma'_{i,j}-\sigma_{i,j}) & 
\quad{\rm if~}(i,j) {\rm ~is~even} \\
\sigma'_{i-1,j}(\sigma'_{i,j}-\sigma_{i,j}) &
\quad{\rm if~}(i,j) {\rm ~is~odd,}
\end{array}\right.
\end{equation}
where $\omega'=\Omega(\omega)$, 
$\omega=\{\omega_{i,j}=(\sigma_{i,j},\tilde{\sigma}_{i,j})\}$, and
$\omega'=\{\omega_{i,j}=(\sigma'_{i,j},\tilde{\sigma}'_{i,j})\}$,
and the identity  
$\Omega^{-1}\circ S^{-1}(\omega)=S^{-1}\circ\Omega(\omega)$,
we obtain
\begin{equation}
J_{i,j,x}(\Omega^{-1}\circ S^{-1}(\omega))=-J_{i,j+1,x}(\omega)
\label{JS}
\end{equation}
in both the cases that site $(i,j)$ is even or odd.
Combining Eqs.\ (\ref{JE2}) and (\ref{JS}), we have
\begin{equation}
\langle J_{i,j,x}E_{l,m}\rangle=-\langle J_{i,j+1,x}E^1_{l,m+1}\rangle.
\end{equation}
Similarly we have 
$\langle J_{i,j,x}\rangle=-\langle J_{i,j+1,x}\rangle$.
Inserting these equalities into Eq.\ (\ref{JDEV}), we arrive at
\begin{eqnarray}
\left.\frac{\partial}{\partial\nabla\beta}
\langle J_{{\rm tot},x}\rangle_{\rm le}\right|_{\nabla\beta=0}
&=&
-\sum_{i,j}\sum_{l,m}l \langle J_{i,j,x}E_{l,m}\rangle \nonumber\\
&=&  
\sum_{i,j}\sum_{l,m}l \langle J_{i,j,x}E_{l,m}^1\rangle \nonumber\\
&=&
\frac{1}{2}\sum_{i,j}\sum_{l,m}l
 \langle J_{i,j,x}(E_{l,m}^1-E_{l,m})\rangle \nonumber \\
&=&
\frac{1}{2}\sum_{i,j}\sum_{l,m}l
\langle J_{i,j,x}
(J_{l,m,x}-J_{l+1,m,x}+J_{l,m,y}-J_{l,m+1,y})\rangle \nonumber \\
&=&
\frac{1}{2}\sum_{i,j}\sum_{l,m}
l\langle J_{i,j,x}(J_{l,m,x}-J_{l+1,m,x})\rangle \nonumber \\
&=&
\frac{1}{2}\sum_{i,j}\sum_{l,m}
\langle J_{i,j,x}J_{l,m,x}\rangle \nonumber \\
&=&
\frac{1}{2}\langle (J_{{\rm tot},x})^2 \rangle .
\end{eqnarray}
This is equivalent to the formula (\ref{C00}).

\begin{figure}
\caption{The structure of the system. The two lines at the edges are 
assigned for heat reservoirs.}
\end{figure}

\begin{figure}
\caption{(a) Distribution of local kinetic energies in Case A. 
Probability that kinetic energy at a site with 
horizontal position $i$ takes 
a value is plotted against the value.  
Calculations were done in a system of size $100\times 100$.\\
(b) Distribution of kinetic energy in Case B obtained 
in the same manner as (a). 
}
\end{figure}

\begin{figure}
\caption{
(a) Scaled temperature profiles in Case A with various system sizes:
$30 \times 30$, $200 \times 200$, and $300 \times 300$. \\
(b) Scaled temperature profiles for Case B.
}
\end{figure}

\begin{figure}
\caption{$\overline{J_{{\rm tot},x}}/L$ measured in the system 
of size $L\times L$ for various boundary temperatures.  
The numbers in the figure $a$-$b$ means that $T_L=a$ and
$T_R=b$.}
\end{figure}

\begin{figure}
\caption{
The partial Kubo sum $\kappa^t$ at various temperatures.
}
\end{figure}

\begin{figure}
\caption{
Themal conductivity $\kappa(T)$ measured in the direct simulation and that
calculated via the Kubo formula in the system of size $200\times 200$.}
\end{figure}

\begin{figure}
\caption{
Thermal conductivity near the critical temperature  
in the systems with different sizes. 
}
\end{figure}

\begin{figure}
\caption{
Numerically computed $\kappa^0(T)$ and 
the mean-field results.  
}
\end{figure}

\begin{figure}
\caption{
Thermal conductivity and $3.5\times\kappa^0(T)$. 
Both are numerically obtained. 
}
\end{figure}

\begin{figure}
\caption{Profile of the local energies in the Q2R of various sizes.}
\end{figure}

\begin{figure}
\caption{
System size dependence of the total energy flux in the Q2R.}
\end{figure}

\begin{figure}
\caption{
Thermal conductivity near the critical point in the Q2R 
computed via the Kubo formula in the system of size $100\times 100$.}
\end{figure}

\epsfxsize=15cm \epsfbox{Fig1.eps}
\vspace*{3cm}
\begin{center}
{\rm \LARGE Fig.1,  K. Saito, S. Takesue, S. Miyashita}
\end{center}

\newpage
\begin{center}
\epsfxsize=15cm \epsfbox{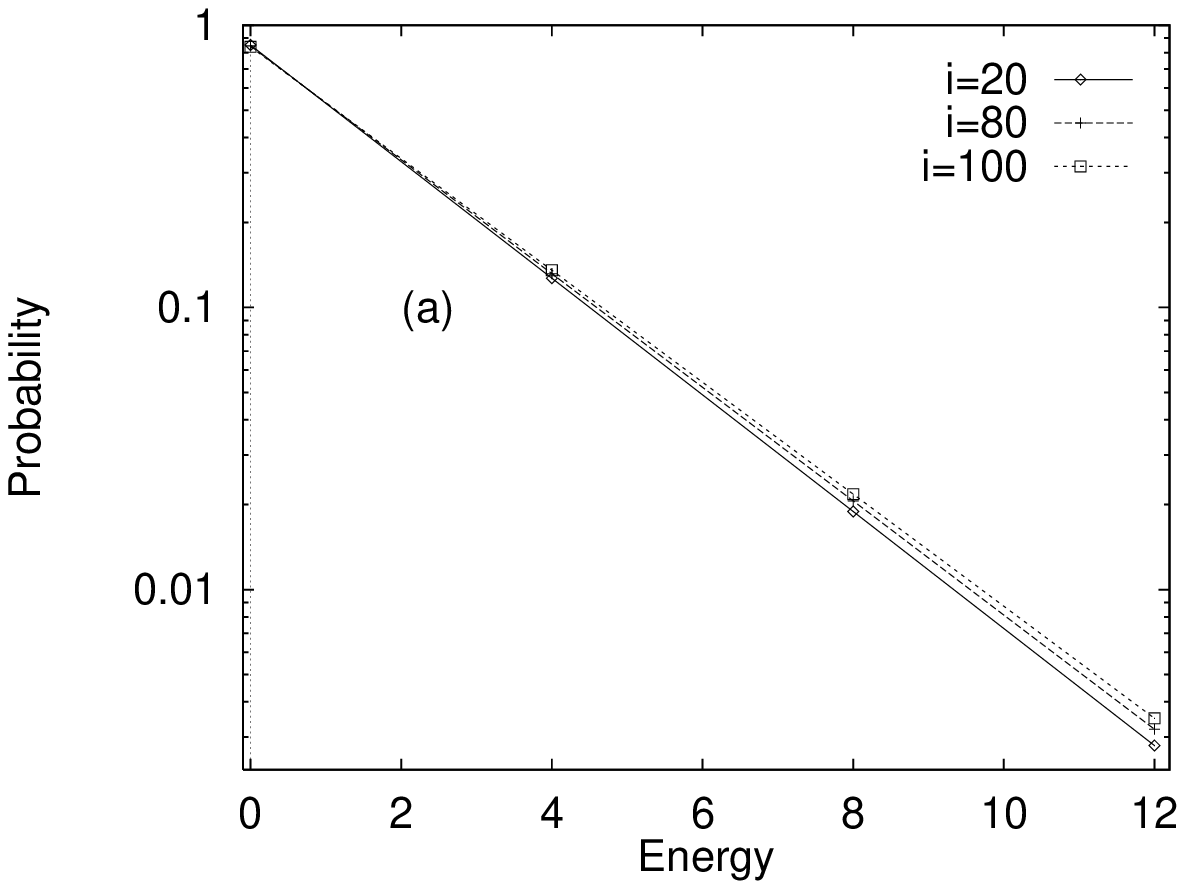}
\end{center}

\vspace*{5cm}
\begin{center}
{\rm \LARGE Fig.2a,  K. Saito, S. Takesue, S. Miyashita}
\end{center}

\newpage
\begin{center}
\epsfxsize=15cm \epsfbox{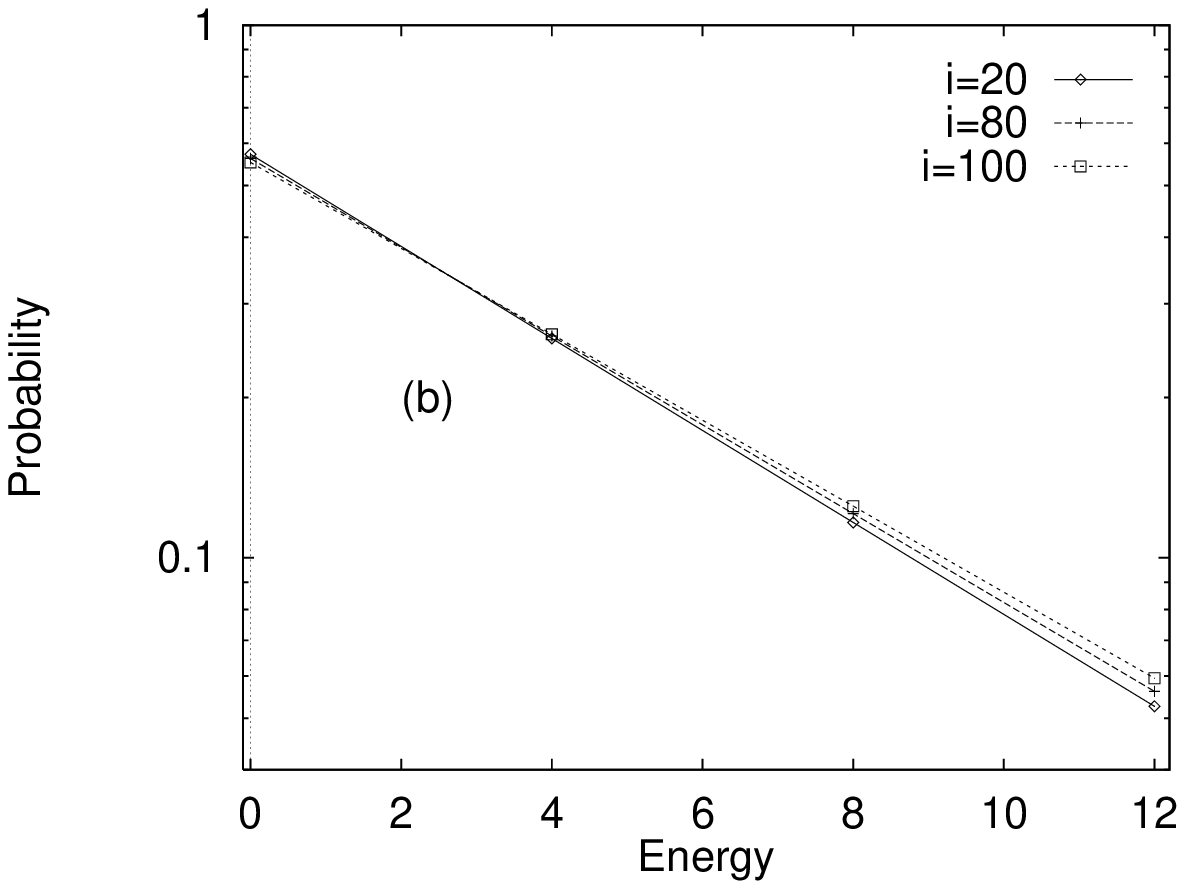}
\end{center}
\vspace*{5cm}
\begin{center}
{\rm \LARGE Fig.2b,  K. Saito, S. Takesue, S. Miyashita}
\end{center}

\newpage
\begin{center}
\epsfxsize=15cm \epsfbox{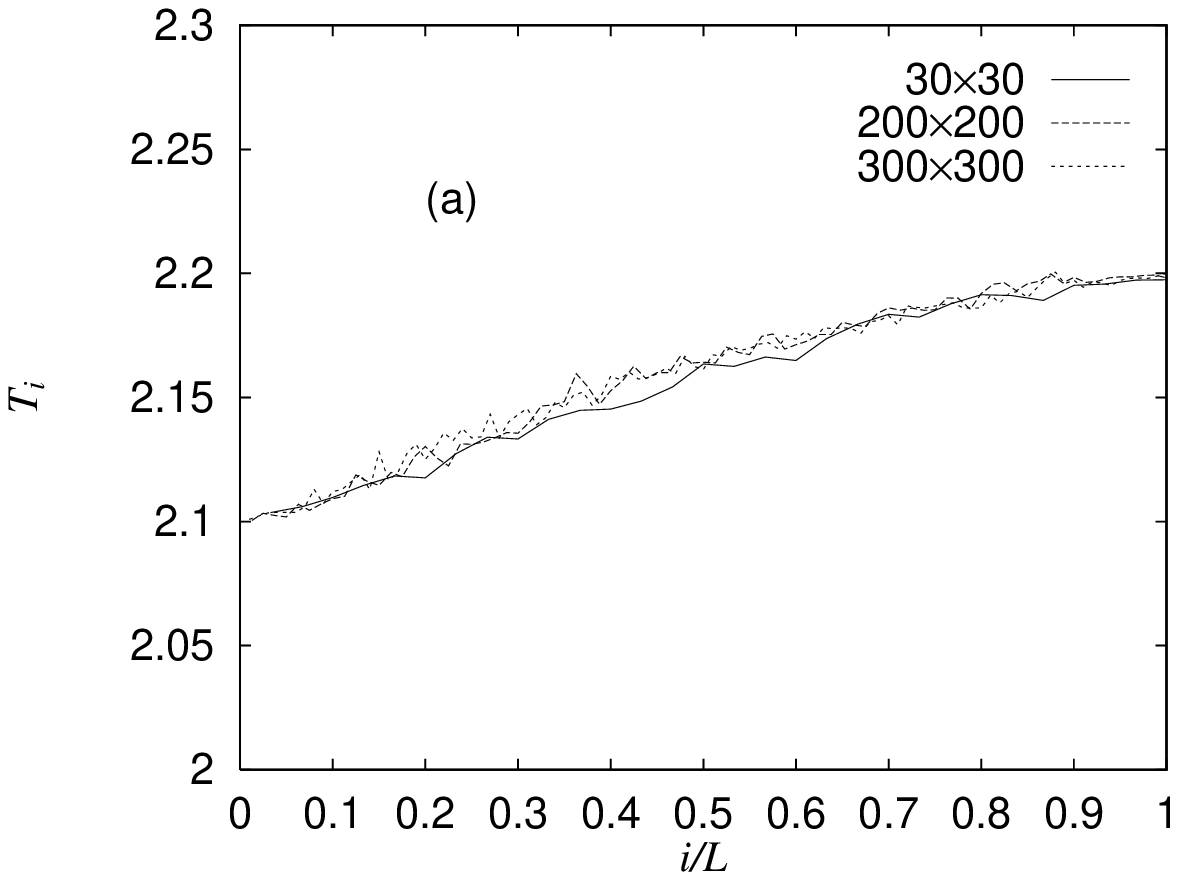}
\end{center}
\vspace*{5cm}
\begin{center}
{\rm \LARGE Fig.\ 3a,  K. Saito, S. Takesue, S. Miyashita}
\end{center}

\newpage
\begin{center}
\epsfxsize=15cm \epsfbox{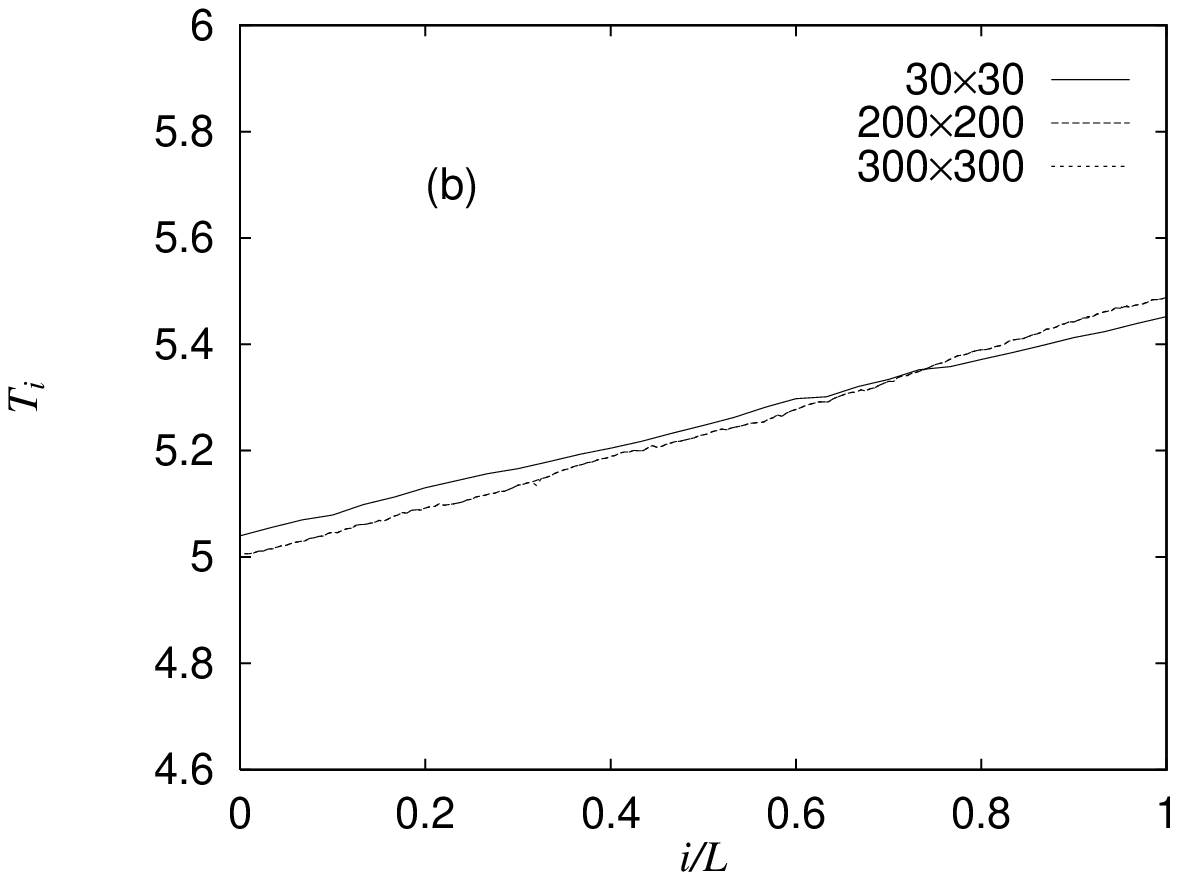}
\end{center}
\vspace*{5cm}
\begin{center}
{\rm \LARGE Fig.\ 3b,  K. Saito, S. Takesue, S. Miyashita}
\end{center}

\newpage
\begin{center}
\epsfxsize=15cm \epsfbox{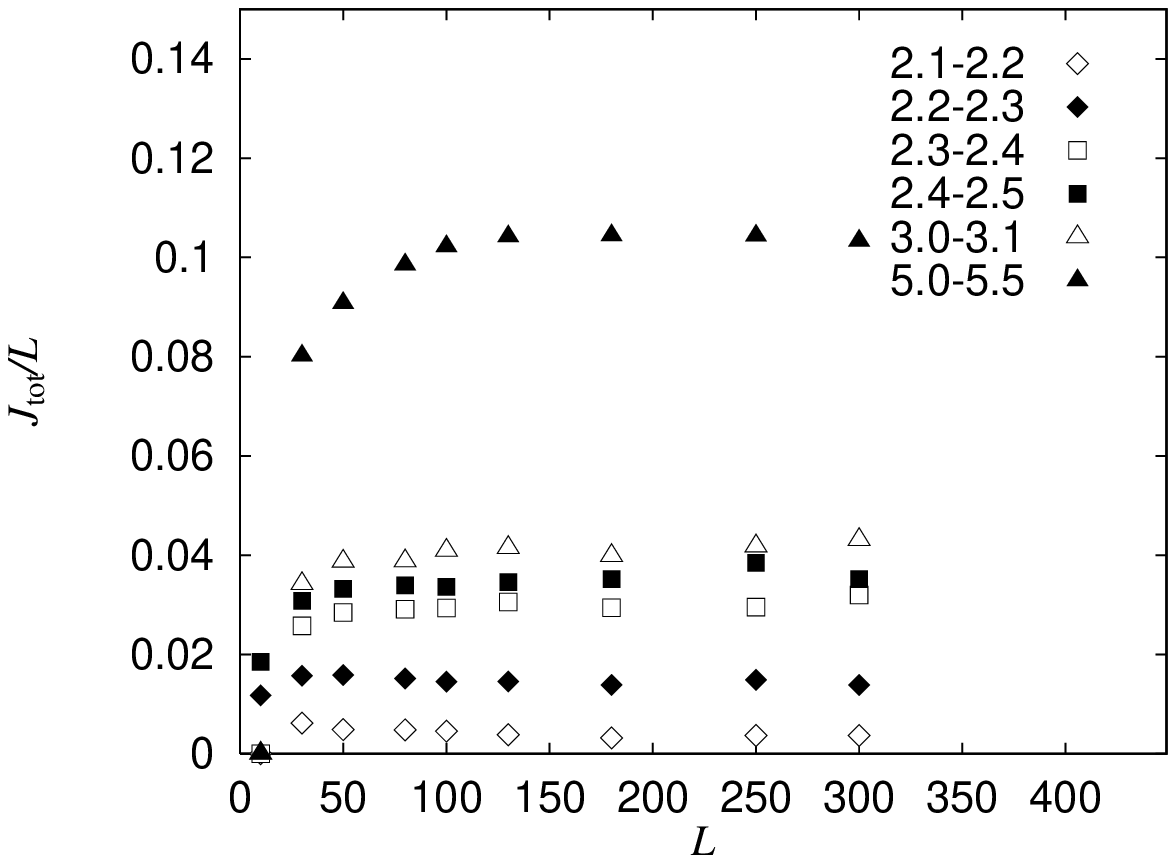}
\end{center}
\vspace*{5cm}
\begin{center}
{\rm \LARGE Fig.\ 4,  K. Saito, S. Takesue, S. Miyashita}
\end{center}

\newpage
\begin{center}
\epsfxsize=15cm \epsfbox{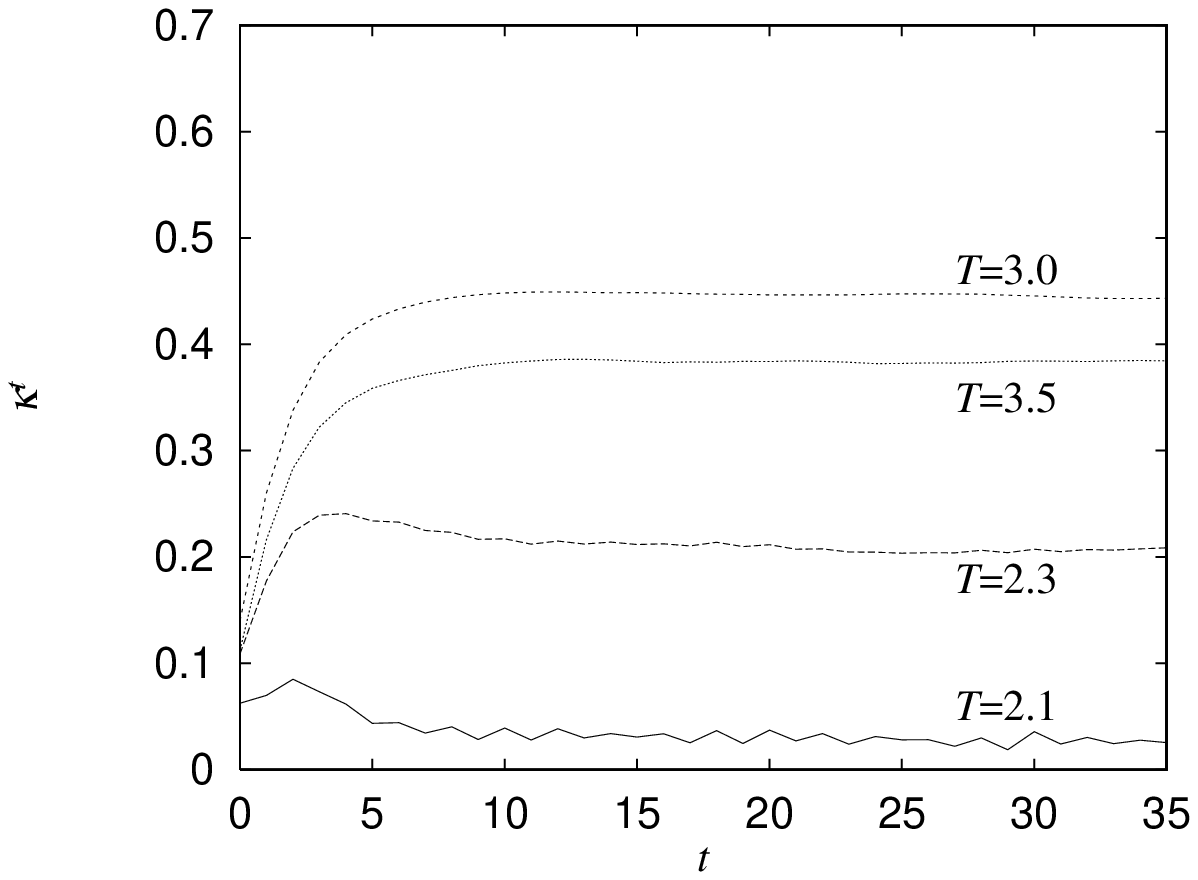}
\end{center}
\vspace*{5cm}
\begin{center}
{\rm \LARGE Fig.\ 5,  K. Saito, S. Takesue, S. Miyashita}
\end{center}

\newpage
\begin{center}
\epsfxsize=15cm \epsfbox{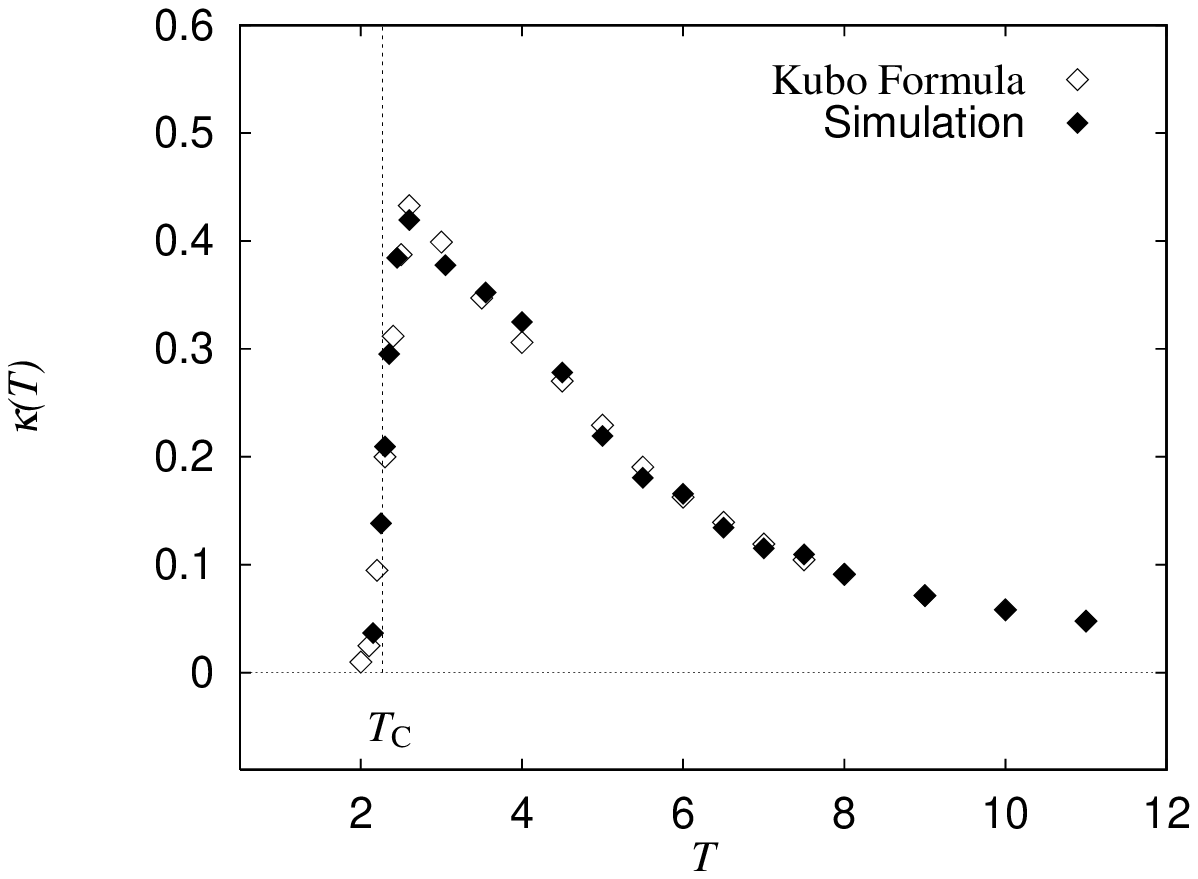}
\end{center}
\vspace*{5cm}
\begin{center}
{\rm \LARGE Fig.\ 6,  K. Saito, S. Takesue, S. Miyashita}
\end{center}

\newpage
\begin{center}
\epsfxsize=15cm \epsfbox{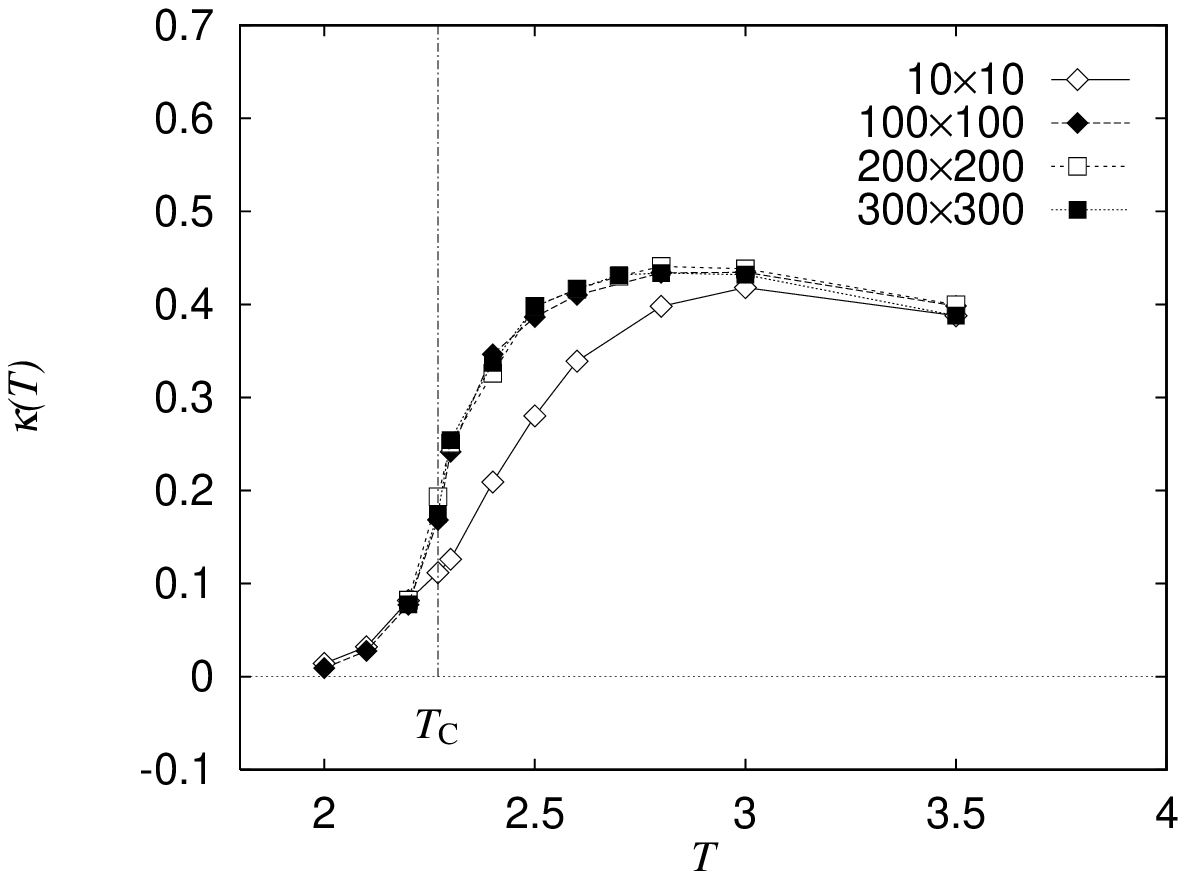}
\end{center}
\vspace*{5cm}
\begin{center}
{\rm \LARGE Fig.\ 7,  K. Saito, S. Takesue, S. Miyashita}
\end{center}

\newpage
\begin{center}
\epsfxsize=15cm \epsfbox{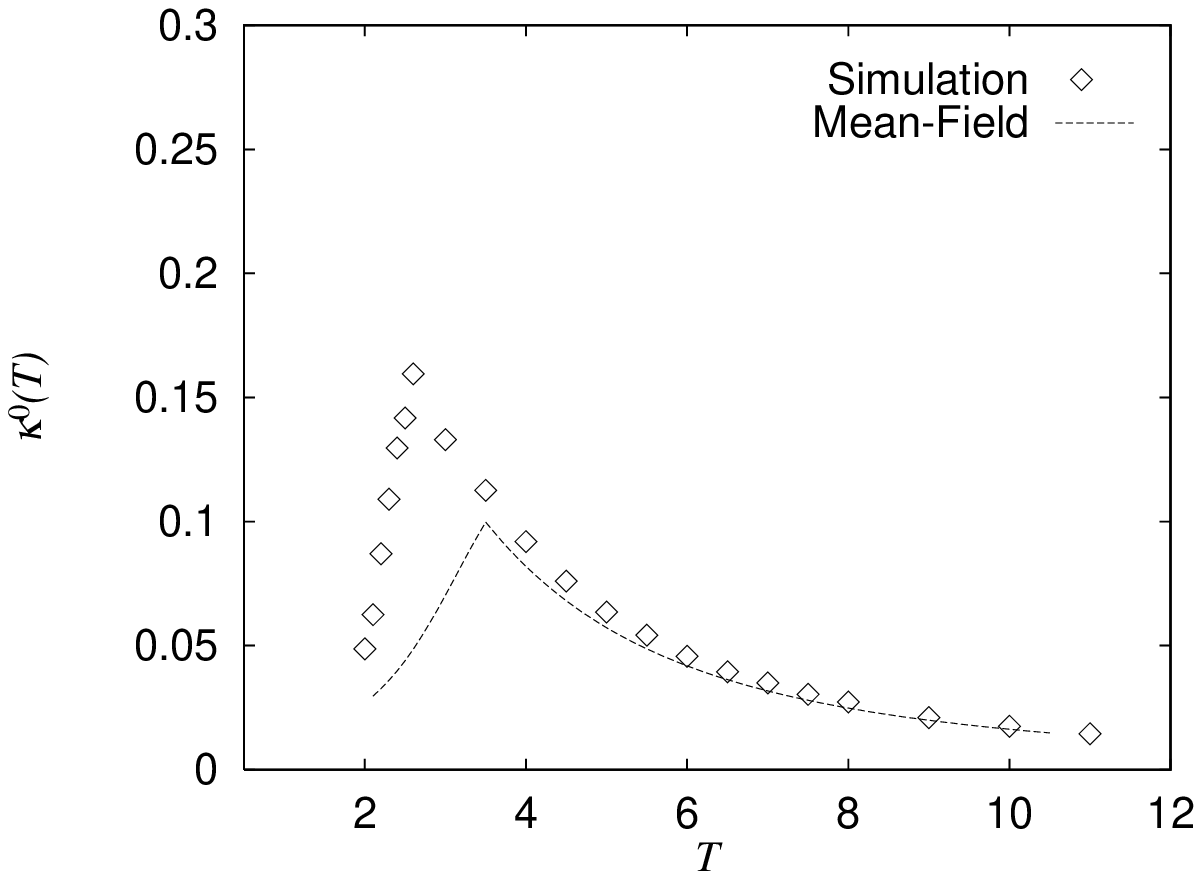}
\end{center}
\vspace*{5cm}
\begin{center}
{\rm \LARGE Fig.\ 8,  K. Saito, S. Takesue, S. Miyashita}
\end{center}

\newpage
\begin{center}
\epsfxsize=15cm \epsfbox{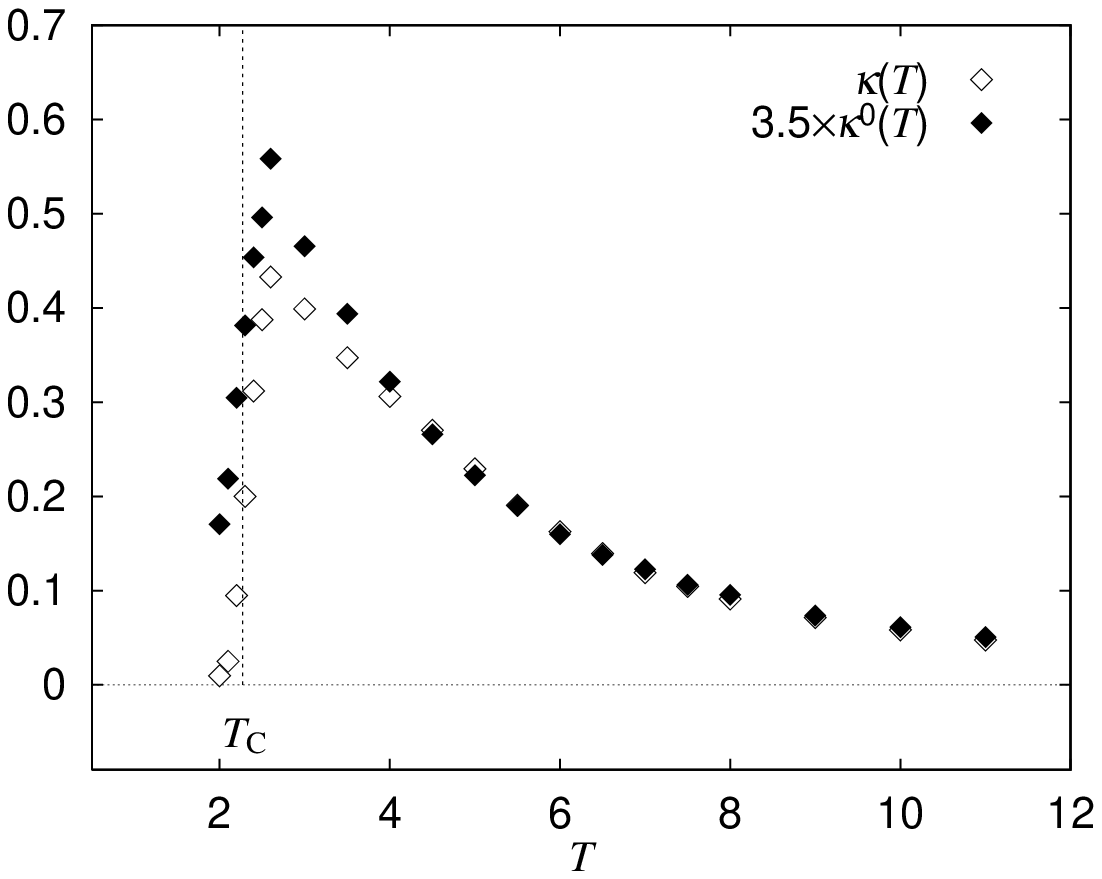}
\end{center}
\vspace*{5cm}
\begin{center}
{\rm \LARGE Fig.\ 9,  K. Saito, S. Takesue, S. Miyashita}
\end{center}

\newpage
\begin{center}
\epsfxsize=15cm \epsfbox{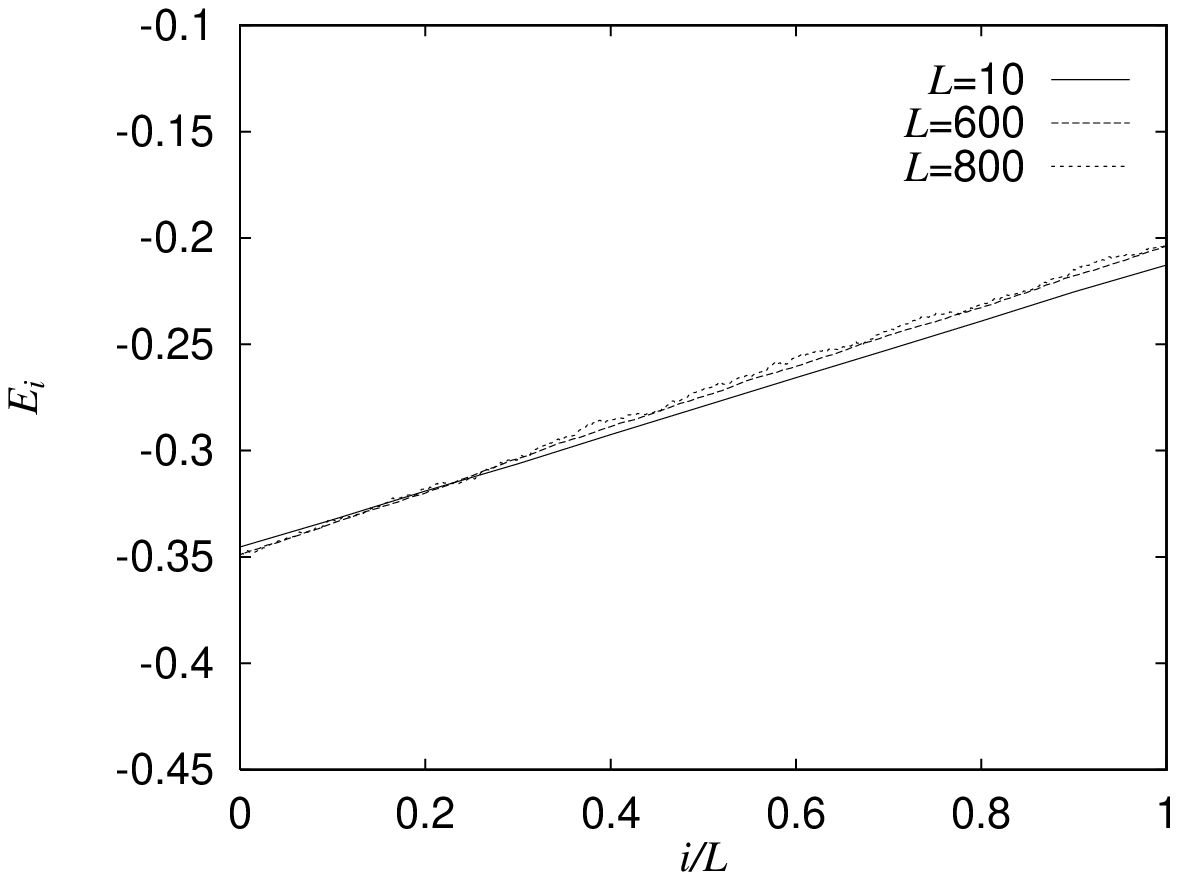}
\end{center}
\vspace*{5cm}
\begin{center}
{\rm \LARGE Fig.\ 10,  K. Saito, S. Takesue, S. Miyashita}
\end{center}

\newpage
\begin{center}
\epsfxsize=15cm \epsfbox{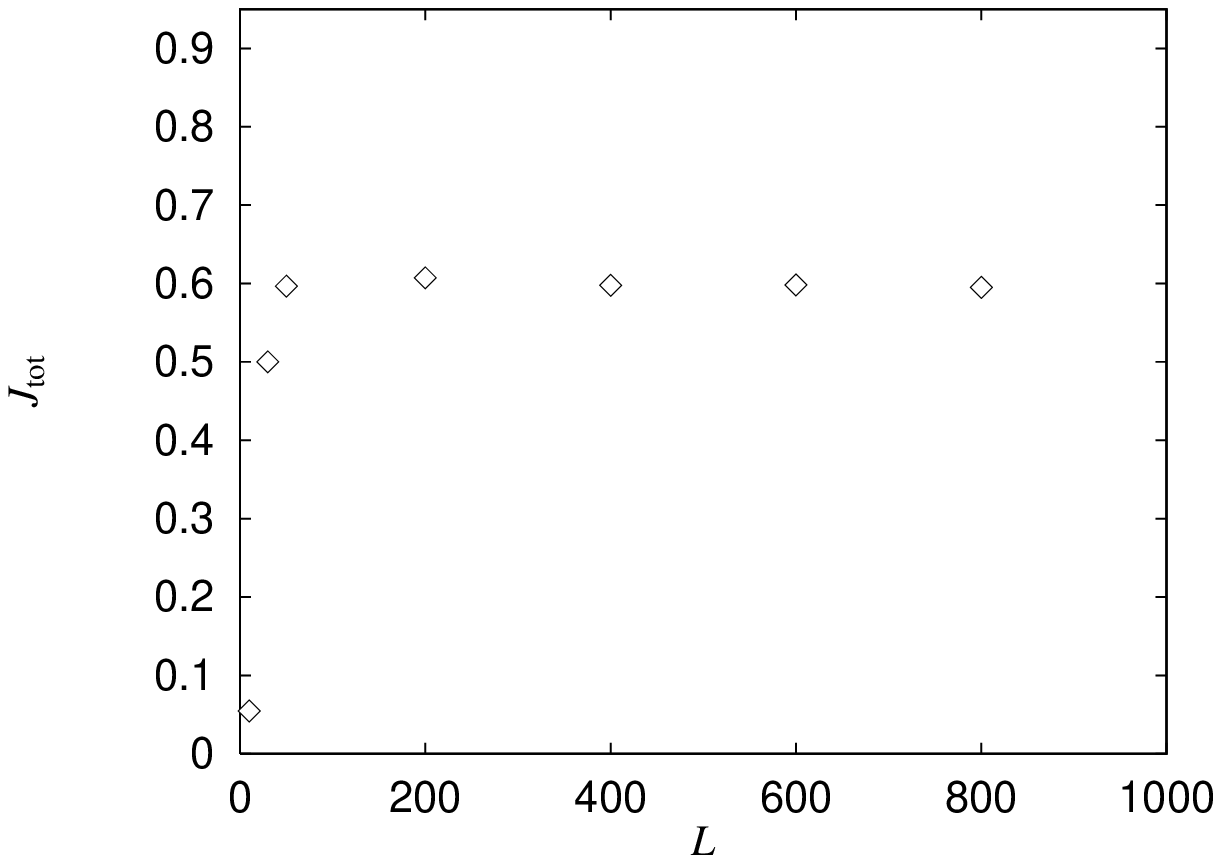}
\end{center}
\vspace*{5cm}
\begin{center}
{\rm \LARGE Fig.\ 11,  K. Saito, S. Takesue, S. Miyashita}
\end{center}

\newpage
\begin{center}
\epsfxsize=15cm \epsfbox{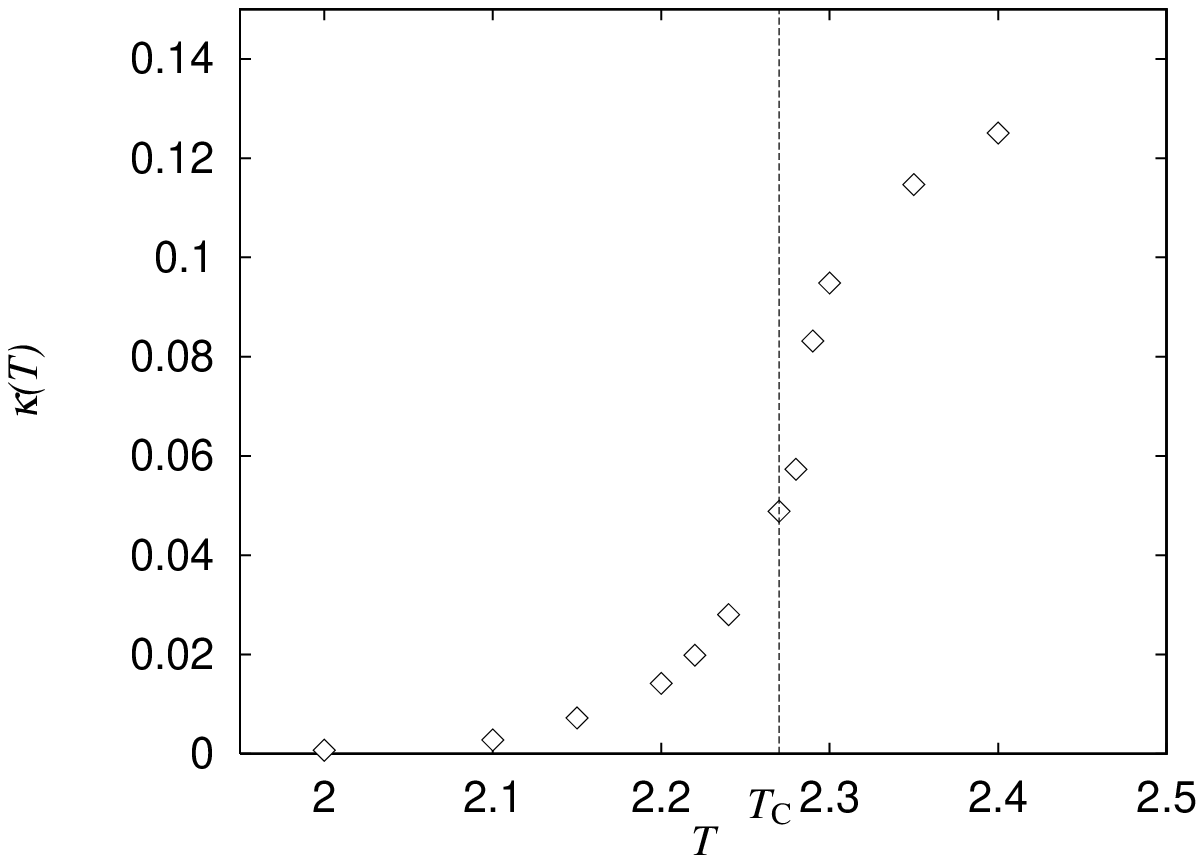}
\end{center}
\vspace*{5cm}
\begin{center}
{\rm \LARGE Fig.\ 12,  K. Saito, S. Takesue, S. Miyashita}
\end{center}

\end{document}